\documentclass[fleqn,usenatbib]{mnras}

\usepackage{newtxtext,newtxmath}

\usepackage[T1]{fontenc}
\usepackage{lipsum}

\DeclareRobustCommand{\VAN}[3]{#2}
\let\VANthebibliography\thebibliography
\def\thebibliography{\DeclareRobustCommand{\VAN}[3]{##3}\VANthebibliography}

\usepackage{graphicx}	
\usepackage{amsmath}	
\usepackage{xcolor}

\title[Stellar yield variations in GCE]{Assessing stellar yields in Galaxy chemical evolution: observational stellar abundance patterns}

\author[J.N. Liang et al.]{
Jinning Liang,$^{1}$
Eda Gjergo,$^{1,2,3}$\thanks{E-mail: eda.gjergo@gmail.com}
and XiLong Fan$^{1}$
\\
$^{1}$School of Physics and Technology, Wuhan University, Wuhan, Hubei 430072, China\\
$^{2}$School of Astronomy and Space Science, Nanjing University, Nanjing 210093, People's Republic of China\\
$^{3}$Key Laboratory of Modern Astronomy and Astrophysics (Nanjing University), Ministry of Education, Nanjing 210093, People's Republic of China
}

\date{Accepted XXX. Received YYY; in original form ZZZ}

\pubyear{2022}

\begin{document}
\label{firstpage}
\pagerange{\pageref{firstpage}--\pageref{lastpage}}
\maketitle

\begin{abstract}
  One-zone Galactic Chemical Evolution (GCE) models have provided useful insights on a great wealth of average abundance patterns in many environments, especially for the Milky Way and its satellites. However, the scatter of such abundance patterns is still  a  challenging aspect to reproduce. The leading hypothesis is that dynamics is a likely major source of the dispersion. 
  In this work we test another hypothesis, namely that different assumptions on yield modeling may be at play simultaneously. We compare whether the abundance patterns spanned by the models are consistent with those observed in Galactic data.
  First, we test the performance of recent yield tabulations, and we show which of these tabulations best fit Galactic stellar abundances. We then group the models and test if yield combinations match data scatter and standard deviation.
  On a fixed Milky-Way-like parametrization of NuPyCEE, we test a selection of yields for the three dominant yield sets: low-to-intermediate mass stars, massive stars, and Type Ia supernovae. We also include the production of r-process elements by neutron star mergers. We explore the statistical properties spanned by such yields. We identify the differences and commonalities among yield sets. We define criteria that estimate whether an element is in agreement with the data, or if the model overestimates or underestimates it in various redshift bins. 
   {While it is true that yields are a major source of uncertainty in GCE models, the scatter of abundances in stellar spectra cannot be explained by a simple averaging of runs across yield prescriptions.} 
\end{abstract}




\section{Introduction}\label{sec:Introduction}

{Galactic Archaeology, here used interchangeably with Galactic Chemical Evolution (GCE), works on the premise that stellar surfaces encapsulate the composition of their progenitor gas, so that little to no nucleosynthesis reaches stellar surfaces via convection or other means.
While this hypothesis is simplistic for a range of stellar categories, works such as \citet{hawkins20} have shown that most stellar wide binary pairs have indeed an extremely homogeneous composition with a dispersion below 0.02 dex. Therefore, any composition observed in the distribution of stellar abundance patterns should reflect the past enrichment histories experienced by the progenitor gas.}

GCE finds its roots in the landmark reviews by \citet{Burbidge57}, known as the B$^2$FH paper, and by \citet{Cameron57}, written independently. Based on the solar abundances compiled by \citet{suess56}, B$^2$FH described the nucleosynthesis processes that generate a wide array isotopic species, including all the stable ones. Ever since B$^2$FH, the field was successful in solving a number of observational challenges. 

One such achievement was the Tinsley-Wallerstein diagram \citep{wallerstein62, tinsley79}, otherwise known as the [$\alpha$/Fe] vs [Fe/H] diagram. This diagram is particularly sensitive to the rates of Type Ia supernovae (SNIa), a source responsible of at least half of the iron enrichment \citep{matteucci09} ahead of core collapse supernovae (SNCC), and SNCC themselves which are very effective at generating $\alpha$ species\footnote{$\alpha$ species are $^4$He, $^{12}$C, $^{16}$O, $^{20}$Ne, $^{24}$Mg, $^{28}$Si, $^{32}$S, $^{36}$Ar, and $^{40}$Ca.}  \citep{matteucci01, pagel09}. Stars residing in different Galactic regions (specifically for the interests of this paper, the halo, the thick disc and the thin disc) are clustered in specific regions of the Tinsley-Wallerstein diagram, providing insights on their past history: halo stars occupy a flat [$\alpha$/Fe]-enhanced, [Fe/H]-poor region.
The flat [$\alpha$/Fe] trend breaks at the age where SNIa begin to enrich the interstellar medium (ISM), [Fe/H]$\geq- 1$\footnote{The square bracket notation represents logarithmic number densities of a species $A$, relative to species $B$, and normalized to solar values:
$$\mathrm{[A/B]} = 
  \log\left(N_{\mathrm{A}}/N_{\mathrm{B}}\right) - \log\left(N_{\mathrm{A}}/N_{\mathrm{B}}\right)_{\odot}.
$$}, 
 causing a downward [$\alpha$/Fe] slope toward zero (i.e., solar values) with increasing iron abundance. This region after the onset of SNIa is where thick disc stars are clustered. Lastly, thin disc stars tend to occupy a space right below thick disc stars, on a flatter [$\alpha$/Fe] slope. 
 
 A recent investigation of the Tinsley-Wallerstein diagram on APOGEE-DR16 data can be found in \citet{spitoni21}, which uses MCMC methods to fit a multi-zone GCE model. The original GCE models were one-zone, meaning that the galaxy is treated as a point. In one-zone models, the metallicity and star formation rate in a galaxy are homogeneously uniform  as a function of time. These type of models are still useful because of the insights they provide on the essential constituents of galaxy evolution. In fact, more sophisticated models should  integrate back to the global properties that emerge from one-zone models. For a recent extensive review on GCE, refer to \citet{matteucci21}.

Despite the many successes and predictions of GCE models, stellar abundances are scattered across a wide range of values, and the exact origin of this scatter is not yet completely understood for every chemical element. There have been numerous attempts at reproducing these patterns, or at least at constraining the conditions that may give rise to them. Most of the models make the sensible assumption that dynamical effects are at play. Proposed phenomena include stellar migration \citep{schoenrich09, spitoni15, johnson21}, gas radial flow \citep{Andrews17, weinberg17, trueman22, chen22}, the inclusion of a variable IMF \citep{yan19} or of a stochastic star formation rate \citep[SFR,][with a specific application to heavy elements]{Cescutti08i}, or Bayesian approaches \citep{Cote17, rybizki17, spitoni20}.

In this work we explore an alternative route, namely we investigate whether multiple enrichment phenomena as described by  different yield models could coexist in a Milky-Way like one-zone galaxy, and whether  they may contribute to the abundance patterns of a select number of elements.
We follow \cite{Romano10}, where they investigated existing yields in literature, and they evaluated which yield in each enrichment channel best fits Galactic stellar abundance data. 
{In this work, we implement sets of more recent yields compared to \citet{Romano10}, where they already investigated \citet{Karakas10} and \citet{Kobayashi06}. We include also \citet{Limongi18, Nomoto13} for massive stars,  \citet{Cristallo11-AGB, Cristallo15} for AGBs, and \citet{NuGrid} for both mass ranges. } We  overview  the yield mass-fraction, final mass, initial mass, initial metallicities. We re-categorize the yields into 3 groups, according to analogous modeling motivations: hypernovae fraction, massive star initial rotational velocity, and delayed/rapid explosion prescriptions for massive stars. We then run on-zone GCE simulations with the use of the OMEGA module in NuPyCEE \citep{Cote17}, using different yield combinations in each run. We therefore show the resulting simulated abundance ratios as a function of iron abundance. We lastly perform a statistical analysis on the aforementioned 3 groups of models and on the data at different metallicity bins. We explore trends and discrepancies.

This paper is structured as follows: Section \ref{sec:Method} contains a description of the observed stellar abundances (Section \ref{sec:Data}), of the stellar yields considered in this work (Section \ref{sec:yields}), of the parameters chosen to run the NuPyCEE simulations (Section \ref{sec:Simulation}), and of the statistical criteria (Section \ref{sec:statistics_tool}) employed in the Results (Section \ref{sec:Results}). In Section \ref{sec:Results} we also explore the performance of the models in reproducing the abundance patterns of our 12 chosen elements. Our conclusions and discussions can be found in Section \ref{sec:Conclusion}, where the summary plots also appear.

\begin{table*}
\centering
\begin{tabular}{lccccccccccccc}
\hline
Paper                                 & C        & N        & O        & Na       & Mg       & Al       & Ca       & Mn       & Ni       & Zn       & Ba       & Eu       & Region of sampling                  \\ \hline
\citet{Carretta00}   & $\times$ & $\times$ & $\times$ & $\times$ & $\times$ &          &          &          &          &          &          &          & halo, thick and thin disc       \\
\citet{Reddy03}      & $\times$ & $\times$ & $\times$ & $\times$ & $\times$ & $\times$ & $\times$ & $\times$ & $\times$ & $\times$ & $\times$ & $\times$ & thin disc   \\        
\citet{Cayrel04}     & $\times$ & $\times$ & $\times$ & $\times$ & $\times$ & $\times$ & $\times$ & $\times$ & $\times$ & $\times$ &          &          & {{halo}}  \\                    
\citet{Israelian04}  &          & $\times$ & $\times$ &          &          &          &          &          &          &          &          &          & halo              \\
\citet{Bensby05}     &          &          & $\times$ & $\times$ & $\times$ & $\times$ & $\times$ &          & $\times$ & $\times$ & $\times$ & $\times$ & thick and thin disc                \\
\citet{Reddy06}      & $\times$ &          & $\times$ & $\times$ & $\times$ & $\times$ & $\times$ & $\times$ & $\times$ & $\times$ & $\times$ & $\times$ & halo, thick and thin disc               \\
\citet{Bensby14}     &          &          & $\times$ & $\times$ & $\times$ & $\times$ & $\times$ &          & $\times$ & $\times$ & $\times$ &          & thick and thin disc \\
\citet{Roederer14a}  & $\times$ & $\times$ & $\times$ & $\times$ & $\times$ & $\times$ & $\times$ & $\times$ & $\times$ & $\times$ & $\times$ & $\times$ & {{halo}}                               \\
\citet{Zhao16}       & $\times$ &          & $\times$ & $\times$ & $\times$ & $\times$ & $\times$ &          &          &          & $\times$ & $\times$ & {{halo}}, thick and thin disc          \\
\citet{Mashonkina17} &          &          &          & $\times$ & $\times$ & $\times$ & $\times$ &          & $\times$ &          & $\times$ & $\times$ & {{halo}}                               \\ 
\citet{Reggiani17}   &          &          &          & $\times$ & $\times$ & $\times$ & $\times$ & $\times$ & $\times$ & $\times$ & $\times$ &          & {{halo}}                               \\
\hline
\end{tabular}
   \caption{Observational data, in chronological order, of the stellar elemental abundances considered in the present work. The first column indicates the references. In the subsequent 12 columns, the $\times$ marks whether a paper contains that column's element. The elements in order are C, N, O, Na, Mg, Al, Ca, Mn, Ni, Zn, Ba, and Eu. The last column indicates in which region of the Milky Way the sampled stars reside.}
   \label{tab:booktabs}
\end{table*}

\section{Methods} \label{sec:Method}

The Methods section is split among observational data, stellar yields investigated, and theoretical models. Within the theoretical models section, we include both a description of the GCE model and of the statistics criteria employed in the analysis.

\subsection{Observational Data}\label{sec:Data}

\begin{table*}
\centering
\def\arraystretch{1.2}
\resizebox{\textwidth}{!}{
\begin{tabular}{cccccc|c}
\hline
Model & \multicolumn{2}{c}{Stellar Yields} & Group & Z range            & M range                   & Comments                                      \\
      & LIMs               & Massive stars &       &                    &                           &                                               \\ \hline
1     & F.R.U.I.T.Y.  & N13           & A     & [0.0001,0.05]      & $\left[1.3, 40.0\right]$  & $f_{\mbox{HNe}}=0.5$                                    \\
2     & K10                & K06           & A     & [0.0001,0.02]      & $\left[1.0, 40.0\right]$  & $f_{\mbox{HNe}}=0.0$                          \\
3     & K10                & K06           & A     & [0.0001,0.02]      & $\left[1.0, 40.0\right]$  & $f_{\mbox{HNe}}=0.5$                          \\
4     & K10                & K06           & A     & [0.0001,0.02]      & $\left[1.0, 40.0\right]$  & $f_{\mbox{HNe}}=1.0$                          \\
5     & K10                & LC18          & B     & [3.236$\times10^{-5}$,0.02]  & $\left[1.0, 120.0\right]$ & Set R, $V_{rot}=0.0$ km/s \\
6     & K10                & LC18          & B     & [3.236$\times10^{-5}$,0.02] & $\left[1.0, 120.0\right]$ & Set R, $V_{rot}=150.0$ km/s                   \\
7     & K10                & LC18          & B     & [3.236$\times10^{-5}$,0.02] & $\left[1.0, 120.0\right]$ & Set R, $V_{rot}=300.0$ km/s                   \\
8     & K10                & LC18          & B     & [3.236$\times10^{-5}$,0.02] & $\left[1.0, 120.0\right]$ & Set R, $V_{rot, avg}$                         \\
9     & F.R.U.I.T.Y        & NuGrid        & C     & [0.0001,0.02]      & $\left[1.3, 25.0\right]$  &       MESA-only (delay)          \\
10    & NuGrid             & K06           & A     & [0.001,0.02]       & $\left[1.0, 40.0\right]$  &            $f_{\mbox{HNe}}=0.5$                \\
11    & K10             & NuGrid           & C     & [0.0001,0.02]      & $\left[1.0, 25.0\right]$  &       MESA-only (delay)        \\
12    & NuGrid             & NuGrid        & C     & [0.0001,0.02]      & $\left[1.0, 25.0\right]$  & MESA-only (delay)                        \\
13    & NuGrid             & NuGrid        & C     & [0.0001,0.02]      & $\left[1.0, 25.0\right]$  & MESA-only (delay wind pre-exp) \\
14    & NuGrid             & NuGrid        & C     & [0.0001,0.02]      & $\left[1.0, 25.0\right]$  & MESA-only (mix)                \\
15    & NuGrid             & NuGrid        & C     & [0.0001,0.02]      & $\left[1.0, 25.0\right]$  & MESA-only (rapid)              \\
16    & NuGrid             & NuGrid        & C     & [0.0001,0.02]      & $\left[1.0, 25.0\right]$  & MESA-only (classical)                                \\ 
\hline\hline
Index & \multicolumn{3}{c}{SNIa yields}            & \multicolumn{2}{c|}{Z range}                   &               Comments                                \\ \hline
a     & \multicolumn{3}{c}{I99}                    & \multicolumn{2}{c|}{0.02}                      & CDD1                                          \\
b     & \multicolumn{3}{c}{I99}                    & \multicolumn{2}{c|}{0.02}                      & CDD2                                          \\
c     & \multicolumn{3}{c}{I99}                    & \multicolumn{2}{c|}{0.02}                      & W7                                            \\
d     & \multicolumn{3}{c}{Ivo12}                  & \multicolumn{2}{c|}{[0.0002,0.02]}             & Stable                                        \\
e     & \multicolumn{3}{c}{T86}                    & \multicolumn{2}{c|}{0.02}                      &                                               \\ \hline
\end{tabular}}
\caption{Legend for the models (\emph{first column}) adopted in the present work. Each model corresponds to a different  yield variation. Numbers from 1 to 16 indicate combinations of yields coming from {low-to-intermediate mass stars (LIMs)} and massive stars. The indices a to e identify the Type Ia SNe (SNIa) yields paired with the above-mentioned models.
The abbreviations stand for the following: NuGrid
\citep{NuGrid}; F.R.U.I.T.Y.
\citep{Cristallo15}; N13
\citep{Nomoto13}; K06 
\citep{Kobayashi06}; K10 
\citep{Karakas10}; LC18 
\citep{Limongi18}; I99 
\citep{Iwamoto99}; Ivo12 
\citep{Seitenzahl12}; T86 
\citep{Thielemann86}. In the last column, \emph{Comments}, we identify the various assumptions on the massive stellar yield computations. More details on these assumptions can be found in the text.}
\label{tab:yieldstabs}
\end{table*}

In this section, we summarize the observational data that appear in the present work. For our purposes, we have chosen to focus on the following elements: C, N, O, Na, Mg, Al, Ca, Mn, Ni, Zn, Ba, and Eu.  The yield performance on each of these elements will be investigated in Section \ref{sec:Results}. All abundances, whether theoretical or observational, are calibrated to photospheric solar data \citep{Asplund09}.

We consider the data-sets presented in Table \ref{tab:booktabs}. The last column includes the Galactic environments -- namely halo, think disc, thick disc, or solar neighborhood and field stars --  in which the sampled stars reside for each study. Some of these studies sampled stars residing in multiple environments:

\begin{figure*}
\centering
\includegraphics[width=\textwidth]{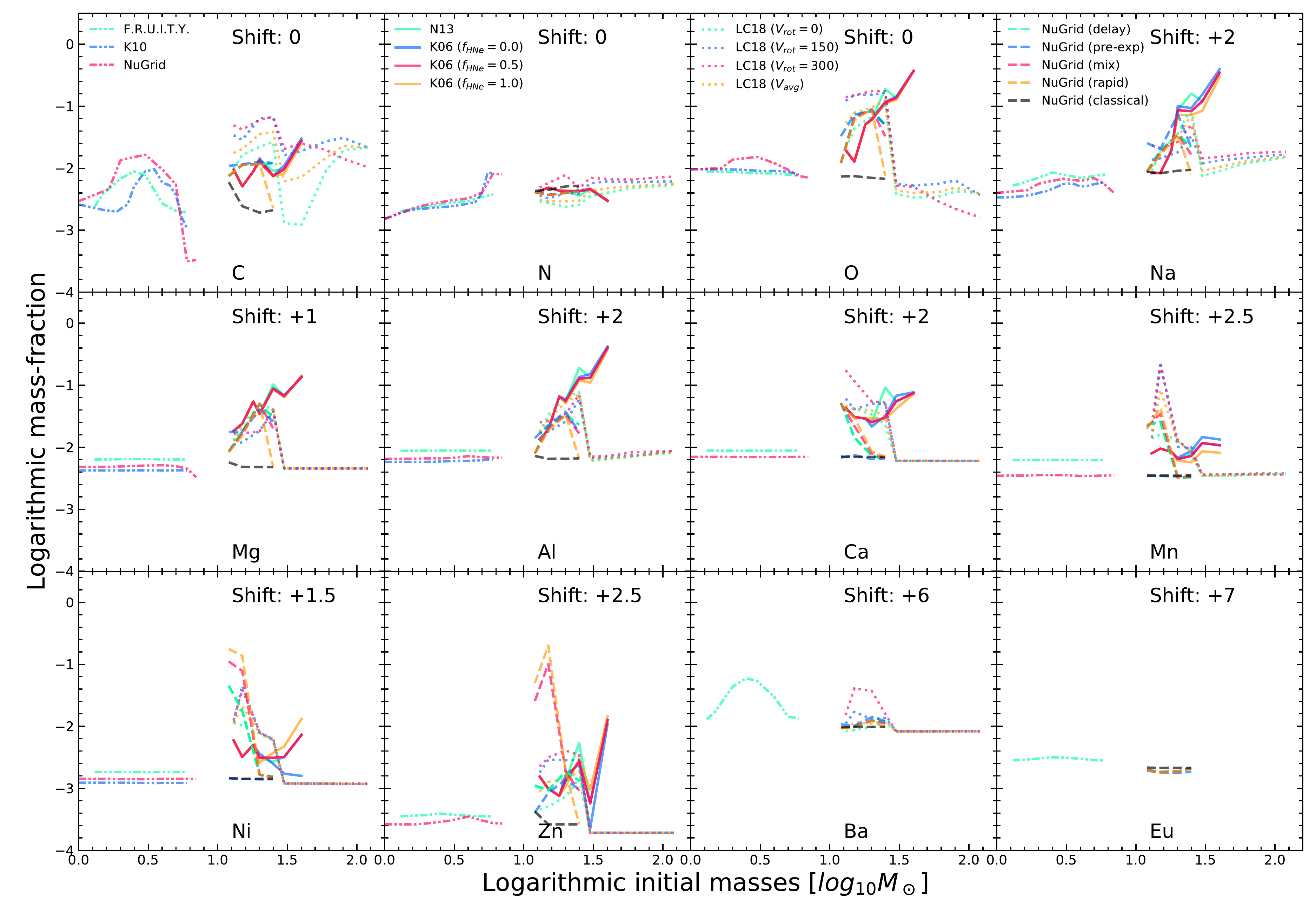}
\caption{Logarithmic mass-fraction (defined as the yield mass over ejecta mass) of the returned yields for 12 elements at solar metallicity, as a function of logarithmic stellar mass. The "shift" label on the upper right indicates, wherever necessary, the factor by which the mass-fractions have been shifted in order to fit the same y-axis range (e.g., the Europium mass-fraction ranges between a value of $10^{-9}$ and $10^{-10}$ while in the plot it ranges between $10^{-2}$ and $10^{-3}$, a shift of +7 dex).
The legends for different stellar yields on the first row follow the description in Table \ref{tab:yieldstabs}. The yields are styled in the following way: all of the low-to-intermediate mass stars (LIMs) yields are represented with the dash-dotted lines; the massive star yields can be represented by solid lines \citep[][]{Nomoto13,Kobayashi06}, by dotted lines \citep{Limongi18}, or by dashed lines \citep{NuGrid}. Among the LIMs yields, the colorings are: green  \citep[F.R.U.I.T.Y., ][]{Cristallo11-AGB,Cristallo15}, included in Models 1 and 9; blue \citep[K10, ][]{Karakas10} included in Models 2 to 8, and black for Model 11; pink \citep[NuGrid, ][]{NuGrid} included in Model 10, 12, 13, 15 and 16, and yellow included in Model 14. 
}
\label{fig:Yieldsvsmass}
\end{figure*}

\begin{figure*}
\centering
\includegraphics[width=\textwidth]{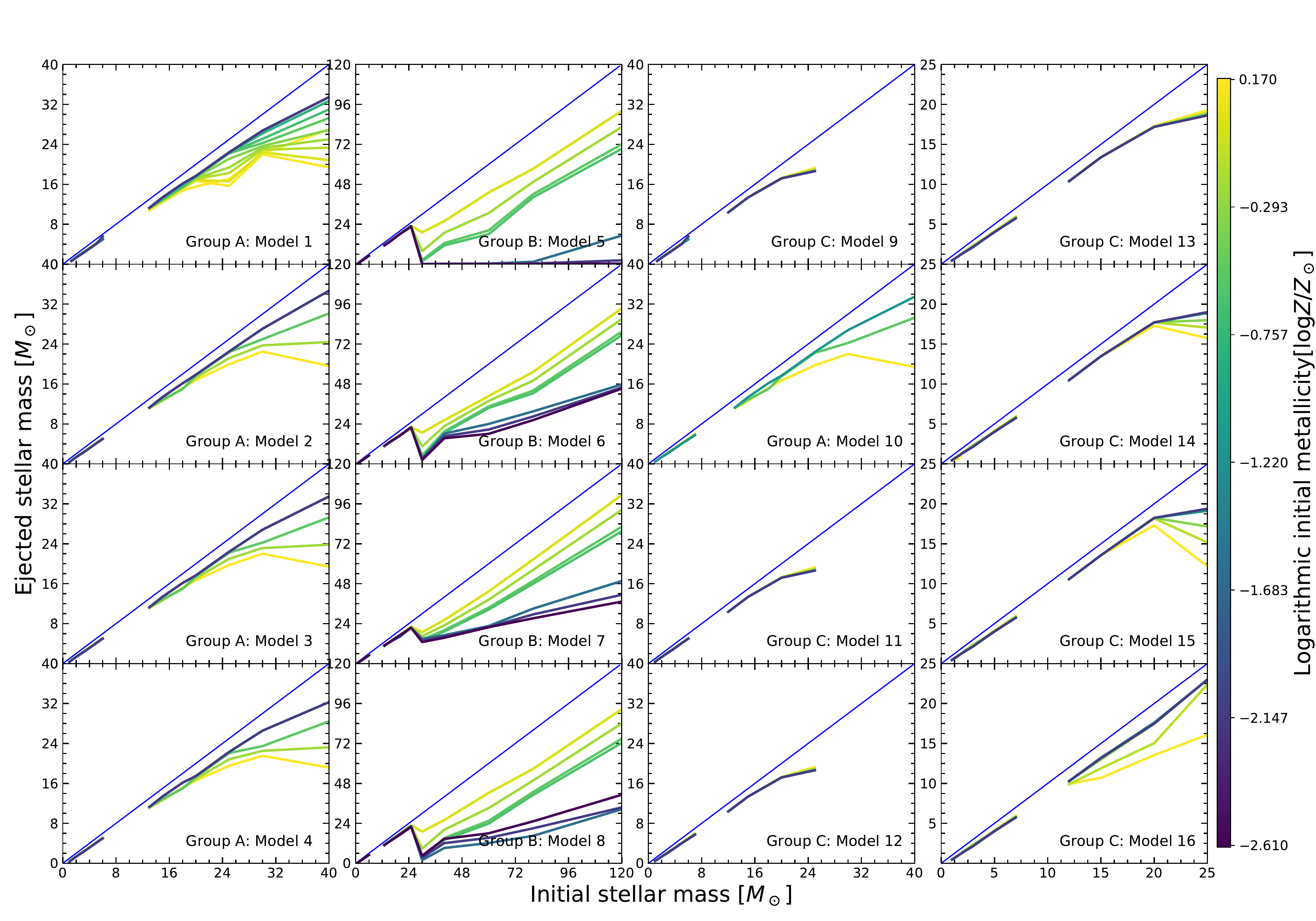}
\caption{The ejected stellar mass versus initial stellar mass for 16 yield tables. The colorbar of the various curves represents different initial metallicities in the range of their yield tables. Here we use a logarithmic scale in the color bar. The lighter the color (yellower), the higher the initial metallicity. The blue line provides a reference, where the ejected
mass equals the initial stellar mass. Groups and Models are described in Table \ref{tab:yieldstabs}. }
\label{fig:finalmass}
\end{figure*}

\begin{enumerate}[i]
    \item Halo stars: 
    \citet{Cayrel04} selected 35 high signal-to-noise ratio giants from the ESO Large Programme "First Stars", where they investigated 70 very metal-poor dwarf and giants whose spectra were captured with the ESO VLT and UVES spectrograph. 17 elemental abundances from C to Zn were reported in their work, i.e. 10 out of the 12 elements we consider.
    \citet{Israelian04}, on the other hand reports N and O only for 31 metal poor stars. Their spectra were measured using multiple instruments. On top of finding a scaling relation for [N/Fe] vs [Fe/H] in the range -3.1 < [Fe/H] < 0, they confirmed the necessity of a primary component to the nitrogen enrichment.
    \citet{Roederer14a} is one of two papers in our sample that reports abundances for all 12 elements. They report, in fact, 48 elemental abundances including an isotopic breakdown for 5 additional species for 191 of the 313 stars observed over the course of 10 years of the Magellan Inamori Kyocera Echelle   campaign. Their efforts focused primarily on the [Fe/H] $\leq$ -2.5 stellar sample.
    \citet{Mashonkina17} also focused on very metal-poor (-4 < [Fe/H] < -2) stars, a selection of 59 stars residing primarily in the halo, but 7 of which were found in dwarf spheroidal galaxies. We chose not to omit these stars. The traced elements are a combination of $\alpha$- and odd-Z-elements, as well as Ni and the main neutron-capture species. They confirm their metal-poor sample is $\alpha$-enhanced by a factor of 2 compared to solar $\alpha$/Fe abundances.
    \citet{Reggiani17} combines literature abundances with their observations to constrain GCE over a wide range of metallicities (-3.6$\leq$ [Fe/H] $\leq$ -0.4). They find their results suggest strong inhomogeneous hypernova enrichment. Group A in our yield set includes variations on this enrichment mechanism. While they include Ba, they don't report CNO elements or Eu.
    \item Thin disc-dominated stars: \citet{Reddy03} is the other reference from Table \ref{tab:booktabs} that investigated all the 12 elements analyzed in our work. They provide abundances for 15 additional elements for 181 F and G dwarf stars, the vast majority \citep[][their Fig. 17]{Reddy03} of which reside in the thin disc. While this work reports stellar abundances from both thin and thick disc, the prevalence of thin disc stars warrants placing them in a separate category.
    \item Thin and thick disc stars: Also \citet{Bensby06} measured chemical abundances in F and G dwarf stars, using the Coud\'{e} Echelle Spectrograph (CES), with 35 and 16 stars residing in the thin and thick disc, respectively. Similarly to \citet{Reddy03}, the thin/thick disc classification was based on kinematic criteria. The work mainly focuses on carbon trends, but measures the abundances of all of our chosen elements, with the exception of N.
    More recently, \citet{Bensby14} again investigated 714 F and G disc dwarf and subgiant stars in the solar neighborhood. This work identifies the bimodality of the [$\alpha$/Fe]-[Fe/H] relation for thin and thick disc stars, the latter being the more $\alpha$-enhanced, despite a significant metallicity overlap in the metal-poor to solar metallicity regime.
    Missing from the \citet{Bensby06, Bensby17} papers are C, N, and Mn and, in the latter, also Eu.
    \item Halo, thick, and thin disc: On a sample of 19 extremely metal-poor stars observed with CES,  \citet{Carretta00} expanded their selection with literature data that measured high signal-to-noise (S/N > 150) spectra for 300 stars spanning a wide range of metallicities. They report abundances for light intermediate elements (C, N, O, Na, and Mg). 
    \citet{Reddy06} then isolated 176 solar-neighborhood F and G dwarf thick-disc candidates. While 95 stars were classified as thick disc stars, 48 could not be classified conclusively, 20 belonged to the halo while 13 to the thin disc. We therefore choose to include this paper among the halo, thick, and thin disc group. Concerning our 12 elements, \citet{Reddy06} provides a complete set with the exception of nitrogen.
    \citet{Zhao16} selects stars with the deliberate aim of collecting spectra for thin, thick disc, and halo stars from $-2.62 \leq$ [Fe/H] $+0.24$ and 17 their elemental abundances from Li to Eu. Of the elements that concern our work, all are present but nitrogen and three species around the iron peak (Mn, Ni, and Zn).
\end{enumerate}

\subsection{Stellar Yields}\label{sec:yields}

In this section, we describe and compare different stellar yields. All of the yields are included in the NuPyCEE database\footnote{\url{ https://github.com/NuGrid/NuPyCEE/tree/master/yield\_tables}}. We define mass-fraction as the ratio of the ejected mass for a specific chemical species divided by the total ejecta mass. A comparison for different stellar yields of 12 elements at solar metallicity is also shown in Fig. \ref{fig:Yieldsvsmass}. The yields we considered in this work and summarized in Table \ref{tab:booktabs} are as follows:
\begin{enumerate}[i]
\item For low and intermediate mass stars, the yields come from:  \cite{Karakas10}; F.R.U.I.T.Y., i.e. \cite{Cristallo11-AGB, Cristallo15}; and NuGrid, i.e. \cite{NuGrid}. 
\item For massive stars, the yields come from: \cite{Nomoto13}, \cite{Kobayashi06}, \cite{Limongi18}, and \citep{NuGrid}.
\item For Type Ia Supernovae, the yields come from: \cite{Iwamoto99}, \cite{Thielemann86}, and \cite{Seitenzahl12}.
\end{enumerate}

Most of the analysis on yield variations in our work concern massive stars. Of the four massive star yield tabulations we consider, two of them \citep{Kobayashi06, Nomoto13} are by the same authors and investigate the impact to the chemical enrichment of the inclusion of hypernova yields in massive stars. Hypernova episodes are set to occur for stars in the mass range from 20 to 40 $M_{\odot}$. The authors define the fraction of stars, $f_{HNe}$, that will experience a hypernova episode. $f_{HNe}$ varies from 0\% (Model 2) to 100\% (Model 4). Three of the models (Model 1, 3, and 10 in Table \ref{tab:booktabs}) are set to $f_{HNe}=0.5$. We call this yield set ``Group A''.

The next set (Models 5 to 8) includes only yield variations in ``Set R'' -- the recommended set -- for \citet{Limongi18}. Models 5, 6, and 7 assume that all of the massive stars rotate at rotational velocities of 0, 150, and 300 km/s, respectively. Model 8 ($V_{rot, avg}$) represents an average of the 3 other rotation velocities. We call this yield set ``Group B''. 

Next come a series of NuGrid models computed with MESA \citep[][Models 9 and 11 to 16]{paxton15}. In the \emph{Comments} column for such models, the term in the parenthesis indicates the remnant mass prescription used for the core-collapse explosion of massive stars as computed in \cite{Fryer12}. ``rapid'' refers to the fast-convection model, i.e. it includes explosions occurring within 0.250 s after bounce; while  ``delayed'' refers to the delayed convection model, which also include later explosions such as the ones dominated by standing accretion shock instability.
``mix'' is a 50/50 mix of the delayed and rapid models, while ``delay wind pre-exp'' adds pre-SN nucleosynthesis and wind nucleosynthesis to the delayed model. Lastly, Model 16 is the classical approach as presented in \cite{Fryer12}. We call this yield set ``Group C''.

For SNIa yields, CDD1, CDD2, W7 are the 3 different models that represent the SNIa detonation scenario from \cite{Iwamoto99}. ``Stable'' for \citet{Seitenzahl12} means that the SNIa yields contain only the stable isotopes. \citet{Thielemann86} reports only one SNIa yield set. The Type Ia SN yields are shown on the bottom of Table \ref{tab:booktabs} with model indices spanning from ``a'' to ``e''.
Table \ref{tab:booktabs} also reports the metallicity and mass range spanned by each yield set. Among SNIa yields, only \citet{Seitenzahl12} reports metallicity-dependent yields in the very metal-poor  vs solar-metallicity regimes.

Each of these references computes yield tabulations on different metallicity grids. The only shared metallicity is $Z=0.02$. In Fig. \ref{fig:Yieldsvsmass} we show the logarithmic mass-fraction (namely the ratio of the yield expressed in solar masses vs. the ejecta mass) as a function of logarithmic masses for all 12 elements and for the various yield prescriptions. Each panel refers to a single element. Except for C, and Ba, LIMs yields have little to no dependence on the stellar mass at these near-solar metallicities. Furthermore, these models are in fairly good agreement with each other. 
We note that the AGB/LIMs yield paper \citet{Karakas10} does not contain yields for  Ca, Mn, Zn, Ba, Eu.

\begin{figure}
\centering
\includegraphics[width=\columnwidth]{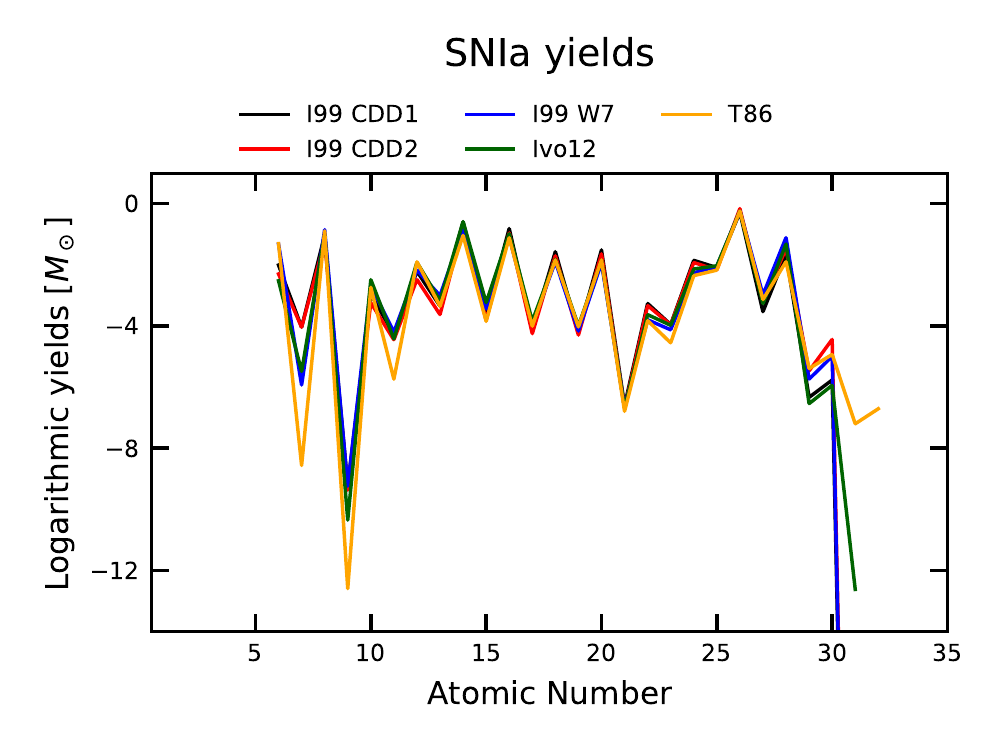}
\caption{Logarithmic yields versus Atomic number for the 5 SNIa yield tables, as indicated in Table \ref{tab:yieldstabs}. The atomic number range goes from 6(C) to 32(Ge)}
\label{fig:SNIayields}
\end{figure}

When it comes to massive stars, there is often a clear dependence of the mass-fraction on the initial stellar mass. We also notice greater variations across models. Therefore, modeling assumptions have a greater impact on the nucleosynthesis of each element. This is particularly evident in the 10-30 $M_{\odot}$ mass range. Only group B \citep{Limongi18} reports yields for stars more massive than 40 $M_\odot$. \cite{Limongi18} accounts for s-process production only, so there is no Eu from this set either.  For the lighter elements, as the atomic number increases, the mass-fraction on average tends to decrease as a function of initial stellar mass -- up to the model's limit for BH formation. The trend inverts as we approach and surpass the iron-peak. 
We point out that the heavy elements Ba and Eu are produced at a particularly low yield. Noting that the mass-fraction is normalized to ejected masses of stars, it is also necessary to show the ejected mass of stellar yields, which we do next. 

In Fig. \ref{fig:finalmass}, we show the ejected mass as function of stellar mass of each model. The color bar represents the logarithmic initial metallicity. 
For massive stars, there is a strong dependence on metallicity when it comes to mass ejecta. 
 Generally, the higher initial stellar mass, the higher ejected mass. But there are some exceptions, for example, Group B show a decreasing trend when the stellar mass is larger than 25 $M_{\odot}$. In Group B \citep[the ][yields, Set R]{Limongi18}, all stars above an initial mass of 25 $M_{\odot}$ collapse into black holes, so the yields above that mass limit come from the ejected wind component. For LIMs, all the models do not show a strong dependence of the ejecta mass on the initial stellar metallicity.  {Models 9, 11, 12, and 13 show little dependence on Metallicity. Model 5, 6, 7, and 8 show that the higher metallicity, the higher the ejected mass, while the trend is opposite for the remaining models}. Please note that the x-axes in Figure \ref{fig:finalmass} are not shared across the various columns, and the maximal stellar mass reached by the yields varies  between 25, 40, and 120 $M_{\odot}$.

In Fig. \ref{fig:SNIayields}, we show logarithmic yields as function of atomic number ranging from 6(C) to 32(Ge) for the 5 SNIa yields we considered. One can see from Table \ref{tab:yieldstabs}, that most SNIa yields have no metallicity dependence, except for Model d that differentiates between a low-metallicity and a solar-metallicity regime. For the majority of the elements that we investigate in the present work, none of the yield tabulations shown in Fig. \ref{fig:SNIayields} vary significantly. The only exceptions are N, Na, and Zn. In the following analysis and predictions we adopt only Model d as the SNIa yield prescription.

\subsection{Theoretical model}\label{sec:theory}

In this section we describe the theoretical basis of both our chemical evolution model and of our analysis. We provide the NuPyCEE prescription employed in the runs which will be analysed in the remainder of this paper. We fix the parametrization for chemical evolution across all runs, and we only vary the yields for Models 1 through 16, as described in Table \ref{tab:booktabs}. Among the enrichment channels, we include: massive stars, LIMs, SNIa, and also neutron star mergers.

\subsubsection{The NuPyCEE Simulation}\label{sec:Simulation}

This work is based on a one-zone chemical evolution prescription where only the yields vary across models. We place particular importance to variations on the yields for massive stars. Our aim is to investigate whether multiple one-zone runs span the same abundance patterns of the data. 
In this study, we exclusively use the chemical evolution code OMEGA \citep[One-zone Model for the Evolution fo GAlaxies][]{Benoit16} implemented within the NuPyCEE\footnote{\url{https://github.com/NuGrid/NuPyCEE/}} framework. A simplified model was introduced in \citet{Cote19} to investigate whether neutron star mergers (NSM) may be the main contributors to the r-process enrichment (exemplified by europium abundances) in the Milky Way. That specific prescription was applied to all the runs in this paper, and will be described in the remainder of this section\footnote{Special thanks to Benoit Cot\'{e} for helping with the setup reconstruction.}.

We keep OMEGA's default initial mass function, the invariant  canonical IMF \citep[][]{Kroupa01}.
\citet{Cote19} simplified the star formation rate (SFR) to a constant value. This assumption is quite typical and representative of rotationally-supported galaxies \citep{schombert19, kroupa20}. The star formation efficiency is set to 0.04. The fraction of the mass ejected into the ISM from simple stellar populations (SSP) is 0.35.

In terms of inflow and outflow, we keep the default I/O Model: the mass mass-loading factor $\eta$   \citep[the ratio between outflow rate and SFR rate][]{Martin99, Veilleux05} is non-zero and equal to 0.5, while the ratio between the inflow rate and the outflow rate is 1.0. 
NuPyCEE treats the evolution of dark matter(DM), which has an impact on the mass-loading factor $\eta$. So the reported value of $\eta$ is its value at the final age of the Galaxy. We also have the relation $\eta\propto v_{out}^\gamma$, where $v_{out}$ refers to the velocity of the outflowing material \citep{Murray05}, which is proportional to the rotation velocity of galaxy. This velocity is also proportional to the virial velocity. This implies that $\eta$ is related to the evolving redshift and virial mass: $\eta(t)=C_\eta M_{vir}(t)^{-\gamma/3}(1+z)^{-\gamma/2}$, where $C_{\eta}$ is a normalization constant calculated in the simulation. We set the value of $\gamma$ equal to 1.0. 

For SNIa,   the number of events per stellar mass formed in every SSP is set to $1.2\times10^{-3}$ [event/$M_{\odot}$], while for NSM, it is set to $3.0\times10^{-5}$. Furthermore, we set the mass ejecta per NSM event to $2.5\times10^{-2} M_{\odot}$. Both SNIa and NSM follow in OMEGA a power law $\propto t^{-1}$ for the delayed time distribution of the detonation or coalescence events, respectively.  For NSM, the coalescence range is set from  10 Myr to 10 Gyr. These are the same parameters adopted by the authors in  \cite{Cote19}. Any other parameter was left unaltered to the default value as provided in the NuPyCEE repository. In the present work, the only parameters we change are the stellar yield tabulations.

\subsubsection{Classification criteria}\label{sec:statistics_tool}
To quantify how well the yield models agree with observational data, we present statistics criteria to investigate median, standard deviation, and scatter.
The criteria of our judgements are defined as:
\begin{subequations} \label{eq:judgement}
    \begin{align}
        J_{M,i,j,k} = &
            \begin{cases}
                \text{Overestimated} &  M_{i,j,k}-M_{\rm dat,j,k}> \delta\\
                \text{In Agreement} & \left| \, M_{i,j,k}-M_{\rm dat,j,k}\, \right| \leq \delta\\
                \text{Underestimated} &  M_{i,j,k}-M_{\rm dat,j,k}< -\delta\\
            \end{cases} \\ 
        J_{\sigma,i,j,k} = &
            \begin{cases}
                \text{Overestimated} & {\sigma_{i,j,k}}/{\sigma_{\rm dat,j,k}} - 1 >   \delta  \\
                \text{In Agreement} & \left| \, {\sigma_{i,j,k}}/{\sigma_{\rm dat,j,k}} - 1 \, \right| \leq \delta \\
                \text{Underestimated} &  {\sigma_{i,j,k}}/{\sigma_{\rm dat,j,k}} - 1< - \delta \\
            \end{cases} \\
        J_{S,i,j,k} = &
            \begin{cases}\label{eq:1c}
                \text{In Disagreement} & {\mathscr{I}_{i,j,k}}/{S_{\rm dat,j,k}}=0\\
                \text{Overestimated} & {\mathscr{E}_{i,j,k}}/{S_{\rm dat,j,k}} > \delta \\
                \text{In Agreement} & ({\mathscr{E}_{i,j,k}}/{S_{\rm dat,j,k}}\leq\delta \\ 
                    & \cap \quad {\mathscr{I}_{i,j,k}}/{S_{\rm dat,j,k}} > 1-\delta)\\
                \text{Underestimated} & ({\mathscr{E}_{i,j,k}}/{S_{\rm dat,j,k}}\leq\delta \\
                    & \cap \quad {\mathscr{I}_{i,j,k}}/{S_{\rm dat,j,k}}\leq 1-\delta)\\
            \end{cases}
    \end{align}
\end{subequations}

where $J_{l,i,j,k}$ is the judgement. The subscript $l$ stands for either the median $M$, the standard deviation $\sigma$, or the scatter $S$. The subscript $i$ identifies the Group or group combination (A, B, C, AB, AC, BC, ABC) of the various Models with their respective yield variations. The subscript $j$ refers to one of the 12 elements considered in this paper. Lastly, the subscript $k$ identifies one of the four metallicity bins (-4.0 $ \lesssim$ [Fe/H] $\lesssim$ -3.0, -3.0 $ \lesssim$ [Fe/H] $\lesssim$ -2.0, -2.0 $ \lesssim$ [Fe/H] $\lesssim$ -1.0, -1.0 $ \lesssim$ [Fe/H] $\lesssim$ 0.0). $M_{i,j,k},\sigma_{i,j,k},S_{i,j,k}$ indicate the median, standard deviation and scatter of the Group $i$ of the element $j$ in the $k$th bin, respectively. And $M_{\rm dat,j,k},\sigma_{\rm dat,j,k},S_{\rm dat,j,k}$ refer to the median, standard deviation and scatter of data of the element $j$. The scatter is defined in the last paragraph of this section. In Eq. \ref{eq:judgement}, we set $\delta=0.3$. Other nearby values were tested ($\delta=0.3\pm0.05$) without significant alterations to the classification outcome.

\begin{figure}
    \centering
    \includegraphics[width=\columnwidth]{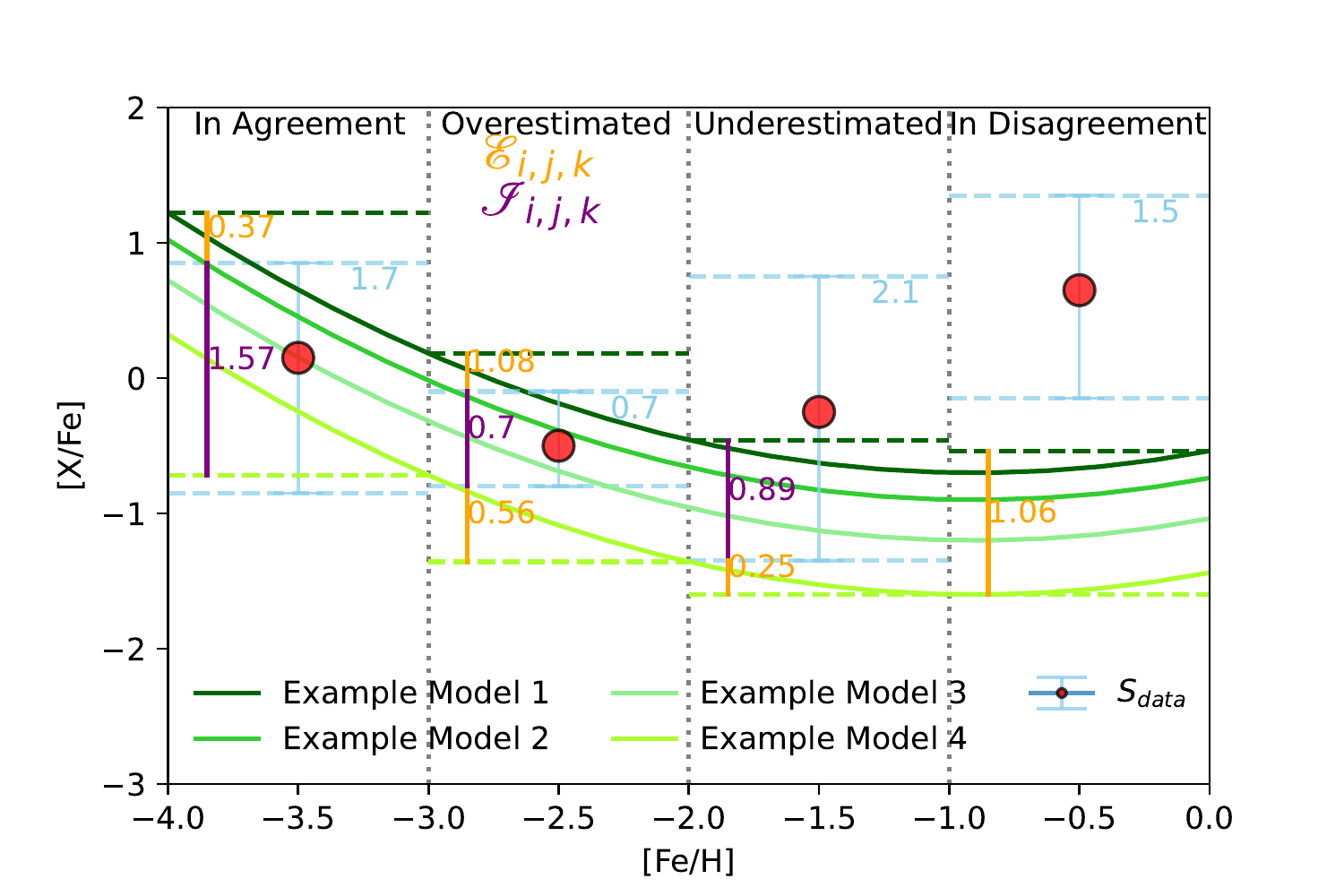}
    \caption{Schematic representation of the judgement for the scatter, as it is introduced in Equation \ref{eq:1c}, with 4 case examples. $\mathscr{I}_{i,j,k}$, the intersection, is shown with the purple vertical segments. $\mathscr{E}_{i,j,k}$, the excess, is shown with the yellow-orange vertical segments. The four green lines are ad-hoc models created as an example for this classification criterion. The red points are the median of the {example} data. As for light blue error bars, they are scatter of the example data. The maximum and minimum of the example models in each of the 4 metallicity bins are shown by the dark green and light green dashed lines, respectively.}
    \label{fig:stat_method}
\end{figure}

In Eq. \ref{eq:1c} $\mathscr{I}_{i,j,k}$ is the intersection of Group $i$ and data for the element $j$ which is the length of the overlap between data and group:
\begin{equation}
 \mathscr{I}_{i,j,k}= \{S_{\rm i,j,k}\} \cap \{S_{\rm dat,j,k}\},
\end{equation}
while $\mathscr{E}_{i,j,k}$ is the excess of Group $i$ and data for the element $j$ which is the length of the non-overlap:
\begin{equation}
 \mathscr{E}_{i,j,k}= \{S_{\rm i,j,k}\} - \{S_{\rm dat,j,k}\}.
\end{equation}

Fig. \ref{fig:stat_method} gives a schematic representation of the judgement for the scatter.  Each metallicity bin displays the value for the $\mathscr{I}_{i,j,k}$ and the excess $\mathscr{E}_{i,j,k}$, as well as the value of the data scatter, $S_{\rm dat,j,k}$, in light blue. Each metallicity bin represents a particular scenario. The lowest metallicity bin (-4 < [Fe/H] < -3) shows the case where the maximum of the example models is larger than the upper light blue error bar -- namely the upper bound of the data scatter --   while the minimum of example models is higher than the lower bound of the scatter. Due to the large intersection and small excess, in this metallicity bin there is an agrement between data and models. In the next metallicity bin (-3 < [Fe/H] < -2), the scatter of the data is contained within the range spanned by the minimum and maximum of the models. In this case, we are overestimating the scatter. The very metal poor metallicity bin (-2 < [Fe/H] < -1) has a large data scatter but low excess. This bin happens to be underestimated with the judgement for the median but also with the judgement for the scatter. In fact, despite the small excess, the scatter intersection is much smaller than the scatter of the data.
In the highest metallicity bin, data and models are in disagreement because there is no intersection.

Theory and models will be more closely in agreement the more each of the judgement criteria tends to the following values: $M_{i,j,k}-M_{\rm dat,j,k}=0$, ${\sigma_{i,j,k}}/{\sigma_{\rm dat,j,k}}=1$ and ${\mathscr{I}_{i,j,k}}/{S_{\rm dat,j,k}}=1$. We therefore decide whether or not a model is in good agreement with the estimate (and similarly, if it overestimates or underestimates the abundance of an element) by the appropriate choice of the arbitrary parameter $\delta$. In the present work, as already stated, $\delta=30$\% accomplishes quite well the task of discriminating the performance of the models.

These statistics criteria employ median, standard deviation and scatter. This information is presented in bin plots and box plots that appear throughout the \ref{sec:Results} Section. For both observational data and groups, the median is defined as the middle among sorte [X/Fe] values in specific metallicity bins for both observational data and for groups. The standard deviation is defined as the amount of variation or dispersion of a set of values in specific metallicity bins, which takes the form of $\sigma=\sqrt{\sum_{i=1}^N(x_i-\Bar{x})/(N-1)}$. Where $x_i$ is the data in specific metallicity bins and $\Bar{x}$ is the mean for these data defined as the sum of data values over the number of data, i.e., $\Bar{x}=\sum_{i=1}^Nx_i/N$. The median and standard deviation of the observational data for each element can be calculated directly by the data we collected from Table \ref{tab:booktabs}. For the models, first we interpolate the curves linearly and then we evaluate the statistical properties in the same metallicity bin. Lastly, we collect the statistical quantities from individual models into groups or group combinations for each metallicity bin.

Some Figures (\ref{fig:CNObox-plot}, \ref{fig:NaMgAlCabox-plot}, \ref{fig:MnNiZnbox-plot}, and \ref{fig:BaEubox-plot}) employ box plots which include 7 quantities: the minimum, the first quartile ($Q1$), the median, the mean, the third quartile ($Q3$), the maximum and the outliers. The difference between $Q3$ and $Q1$ is the inter-quartile range ($IQR$). The maximum is defined as $Q3+1.5IQR$ while minimum is defined as $Q1-1.5IQR$. The data points that reside outside the minimum-maximum range are the outliers.  We define the subtraction of minimum from maximum as the scatter $S_{i,j,k}$ (see Eq. \ref{eq:judgement}). In the box plots of Sec. \ref{sec:Results}, the boxes represent $Q1$ and $Q3$ and use upper arms and lower arms to represent maximum and minimum, respectively. The mean and median are also shown inside the boxes. With that, we can investigate succinctly the statistical features of the data, and we can check whether the models follow similar trends. To compare the spread of the abundance data from stellar spectra and model estimates, we create the pie-charts in  Section \ref{sec:Conclusion} to evaluate the comparisons.

\begin{figure*}
\centering 
    \includegraphics[width=\textwidth]{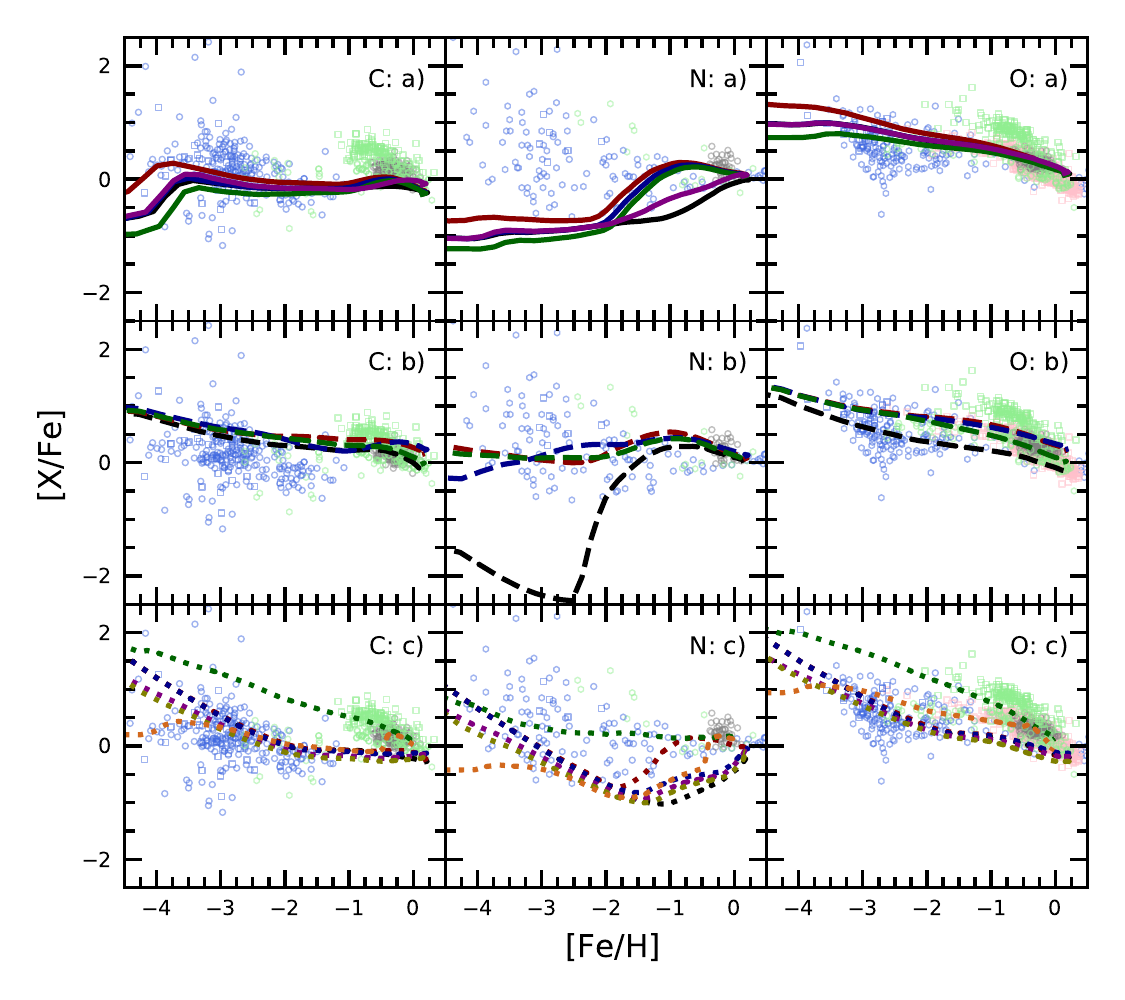}
    \caption{The [X/Fe]-[Fe/H] relation in the Milky Way, where  each column from left to right represents C, N, and O respectively. Each row refers to a different group. The models in the first row belong to Group A and are drawn with a solid line. The models in the middle row belong to Group B and are drawn with a dashed line. The models in the bottom row belong to Group C and are drawn with a dotted line.
    Next we describe the color patterns. \emph{first row:} Model 1 (black), Model 2 (dark red), Model 3 (dark blue), Model 4 (dark green), Model 10 (purple); \emph{second row:}  Model 5 (black), Model 6 (dark red), Model 7  (dark blue), Model 8 (dark green); \emph{third row:}  Model 9 (black), Model 11 (dark red), Model 12 (dark blue), Model 13 (dark green), Model 14 (purple), Model 15 (olive), Model 16  (chocolate). The yield prescriptions of the models are described in Table \ref{tab:yieldstabs}.
    Concerning the color patterns for the observational data: halo stars are shown in royal blue, halo stars as well as thick and thin disc stars are shown in light green, thick and thin disc stars are shown in pink, thin disc-only stars are shown in gray. All of the data are shown with empty markers. Specifically, the observational data sources are displayed as follows: the halo stars in royal blue, the squares come from \citet{Cayrel04}, the hexagons come from \citet{Roederer14a}, while the circles from \citet{Israelian04}. Light green halo and thin-and-thick disc stars are either circles \citep{Zhao16}, squares \citep{Reddy06} or hexagons \citep{Carretta00}. The pink thin and thick disc stars are either circles \citep{Bensby05} or squares \citep{Bensby14}. The gray circles of thin disc stars come from \citet{Reddy03}. 
    }\label{fig:CNOabundance}
\end{figure*}

\begin{figure*}
\centering
    \includegraphics[width=\textwidth]{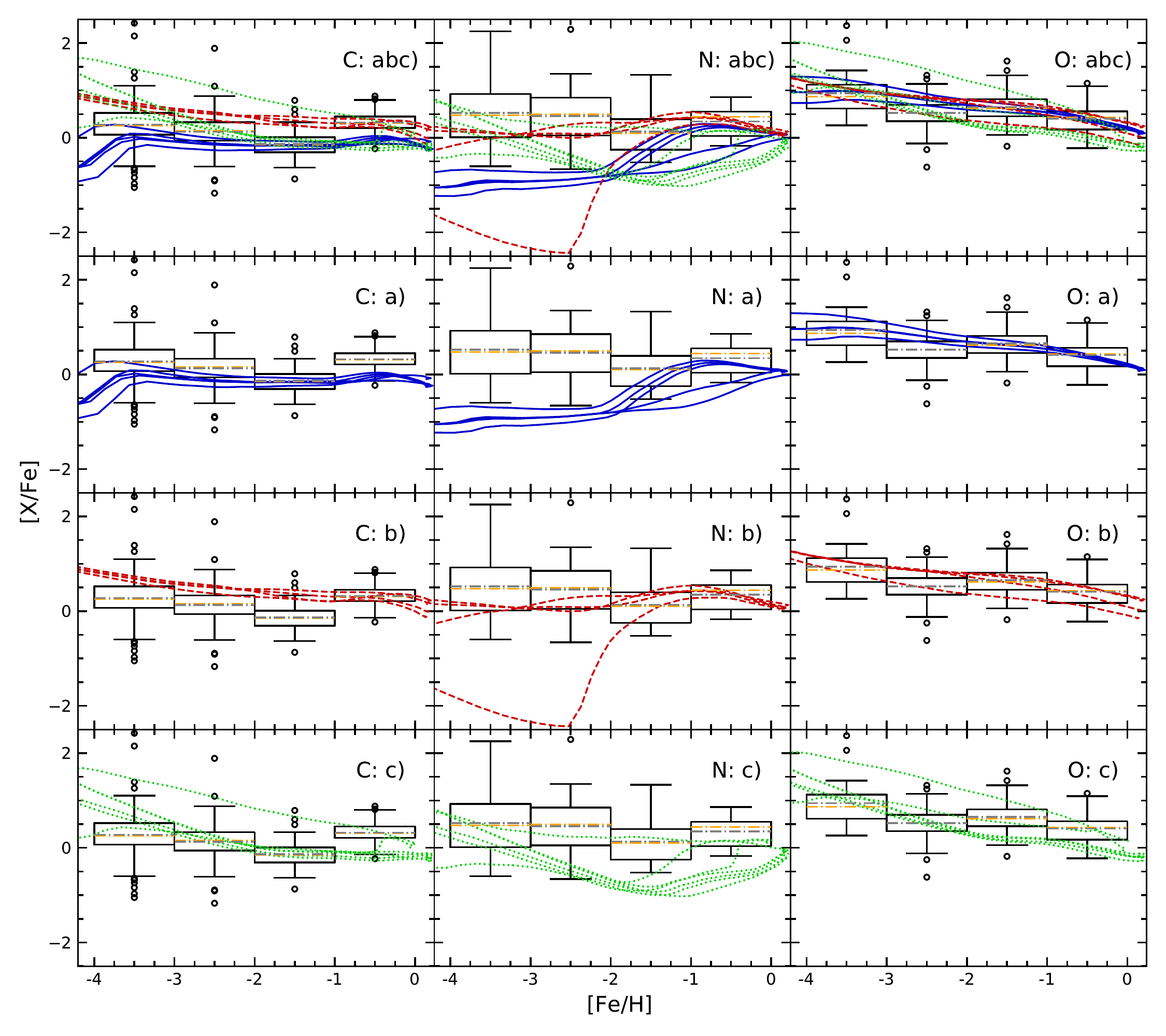}
    \caption{Similarly to the previous box plots, we show the [X/Fe]-[Fe/H] relation for the elements (C, N, O in each column from left to right) and each group, with the addition of Group ABC -- the combination of all three base groups -- on the top row. Unlike Fig. \ref{fig:CNOabundance}, the data are grouped in four metallicity bin in the range -4 < [Fe/H] < 0 and the box plot for each bin is shown in black. The median (dot-dashed orange lines inside the boxes) and average (dot-dashed grey lines inside the boxes) of the data are also shown. All of Group A models are shown in solid blue lines, all of Group B models are shown in dashed red lines, while all of Group C curves are shown in dotted green lines.}
\label{fig:CNObox-plot}
\end{figure*}

\section{Results}\label{sec:Results}
In this section, we investigate the abundance patterns of each element in each of the 16 models as a function of metallicity, using iron as the metallicity indicator.  We then employ the statistics criteria from Section \ref{sec:statistics_tool} to evaluate how closely the model groups match the data.
Since different SNIa yield tables show nearly the same result, we show the simulation with SNIa yields ``d'' \cite{Seitenzahl12}, as labeled in the bottom of Table \ref{tab:yieldstabs}. We recall that we have divided the 16 models into 3 groups: Group A includes Model 1, 2, 3, 4, 10. Group B includes Model 5, 6, 7, 8. Group C includes Model 9, 11, 12, 13, 14, 15, 16. The models in each group normally vary only one parameter or nucleosynthetic prescription at a time -- and primarily among the yields of massive stars. Models in Group A vary the hypernova fraction according to \citet{Kobayashi06}, models in Group B vary the rotational velocity of massive stars according to \citet{Limongi18}, while Group C \citep[NuGrid, ][]{NuGrid} considers either fast- or delayed-convection explosions \citep[][fast-convection  includes explosions occuring 250 ms after "bounce", while delayed convection explosoins extend beyond this limit, including explosions dominated by the standing accretion shock instability]{Fryer12}, as it is applied within the code Modules for Experiments in Stellar Astrophysics \citep[MESA, ][]{Paxton11, Jones15}. We then combine any 2 or 3 groups at a time in Group AB, Group AC, Group BC and Group ABC.

Our results are divided into the following four sections: CNO elements in Section \ref{sec:CNO}, $\alpha$ and odd-Z elements in Section \ref{sec:alpha}, iron-peak elements in Section \ref{sec:iron}, and neutron capture elements in Section \ref{sec:ncapture}. For each of these categories we will present three figures: a first plot that depicts the [X/Fe]-[Fe/H] relation with the data scatter and the models where each row refers to a different group while each column a different element; a second plot, similar to the first plot, in which the data are sorted in 4 metallicity bins and then represented by means of box-plots. In these plots, the top row depicts all 16 models simultaneously; and a third plot, where in the same 4 metallicity bins, the median and standard deviation as defined in Section \ref{sec:statistics_tool} are shown for the various groups and group combinations.

\begin{figure*}
\centering
    \includegraphics[width=\textwidth]{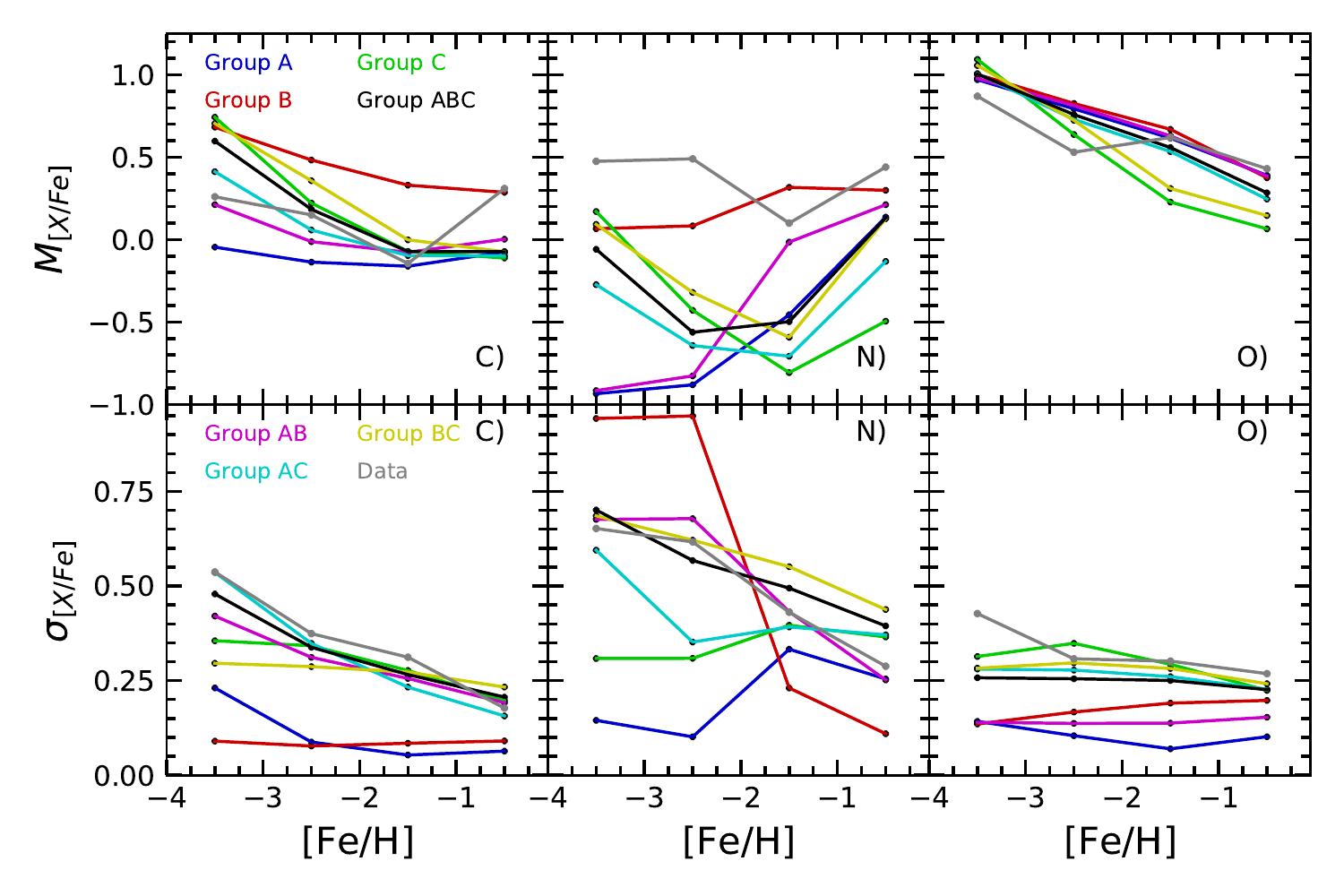}
    \caption{In metallicity bins consistent with the box-plots of Fig. \ref{fig:CNObox-plot} and Fig. (\ref{fig:piechart1}, \ref{fig:piechart2}), the median ($M_{i,j,k}$, \emph{top row}) and standard deviation ($\sigma_{i,j,k}$\emph{bottom row}) for each group and group combination. 
    The black points are shown in the middle of their metallicity bins (-4.0 dex $ \lesssim$ [Fe/H] $\lesssim$ -3.0 dex, -3.0 dex $ \lesssim$ [Fe/H] $\lesssim$ -2.0 dex, -2.0 dex $ \lesssim$ [Fe/H] $\lesssim$ -1.0 dex or -1.0 dex $ \lesssim$ [Fe/H] $\lesssim$ 0.0 dex). All of the panels adopt the same colors and styles: data (grey lines), Group A (blue lines), Group B (red lines), Group C (green lines), the combination of Group A and Group B (Group AB, purple lines), the combination of Group A and Group C (Group AC, cyan lines), the combination of Group B and Group C (Group BC, yellow lines), the combination of Group A and Group B and Group C (Group ABC, black lines). A description of the statistics criteria can be found in Section \ref{sec:statistics_tool}. The data and models used are consistent with the previous figures, Fig. \ref{fig:CNOabundance} and Fig. \ref{fig:CNObox-plot}.}
\label{fig:CNOsta}
\end{figure*}

\subsection{CNO elements}\label{sec:CNO}

The CNO elements are the most abundant atomic species synthesized by stars. Given that a wide range of nuclear processes can yield oxygen, carbon, and nitrogen, their chemical evolution is rich and complex \citep[for a recent review, see ][]{Romano22}. In \citet{Salpeter52}, it was found that the two main species of carbon and oxygen, $^{12}$C and $^{16}$O, as well as $^{18}$O, could be produced during He-burning \citep{Hoyle54, Freer14} by either triple-$\alpha$ or $^{12}$C($\alpha$, $\gamma$)$^{16}$O processes \citep{Clayton03}. Other important isotopes can instead form in the cold \citep[$^{13}$C, $^{14}$N, and $^{17}$O, ][]{Bethe39} and hot \citep[> $10^8$ K $^{15}$N, ][]{Dearborn78}  CNO cycles during H-burning \citep[for a review, see ][]{Wiescher10}. Furthermore, experiments directly tailored to measure key reactions involved with the CNO cycle have been conducted by the LUNA Collaboration \citep{Ananna22}.

Despite these advancements, one of the most significant open questions in stellar nucleosynthesis concerns the Carbon-to-Oxygen abundance ratio at the end of He burning \citep{Buchmann06}. The $^{12}$C($\alpha$, $\gamma$)$^{16}$O cross section in stellar environments are out of reach of current laboratory sensitivities \citep{Kunz02}, so the results need to be extrapolated from experiments performed at higher energies \citep{Aliotta22}. And similarly for the triple-$\alpha$ reaction, \citet{Kibedi20} found that improved measurements lead to a 34\% increase of its reaction rates, with unexplored consequences on nucleosynthetic computations \citep[a study on the impact of nuclear reaction uncertainties be found in ][]{Fields18}.

\subsubsection{Carbon}
Carbon is the second most abundant metal in terms of cosmic as well as gas-phase and stellar Galactic abundances \citep{Nieva12}, right after oxygen. Despite this, it still poses some challenges. The recent consensus is that half of the carbon is synthesized by massive stars, while the remaining half is produced by low-mass (1-4 $M_{\odot}$) AGB stars \citep{Kobayashi11-popIII}. In fact, new computations that consider both rotation and mass loss of massive stars can yield a significant amount of $^{12}$C \citep{Limongi18, Romano20}.

\begin{figure*}
\centering
    \includegraphics[width=\textwidth]{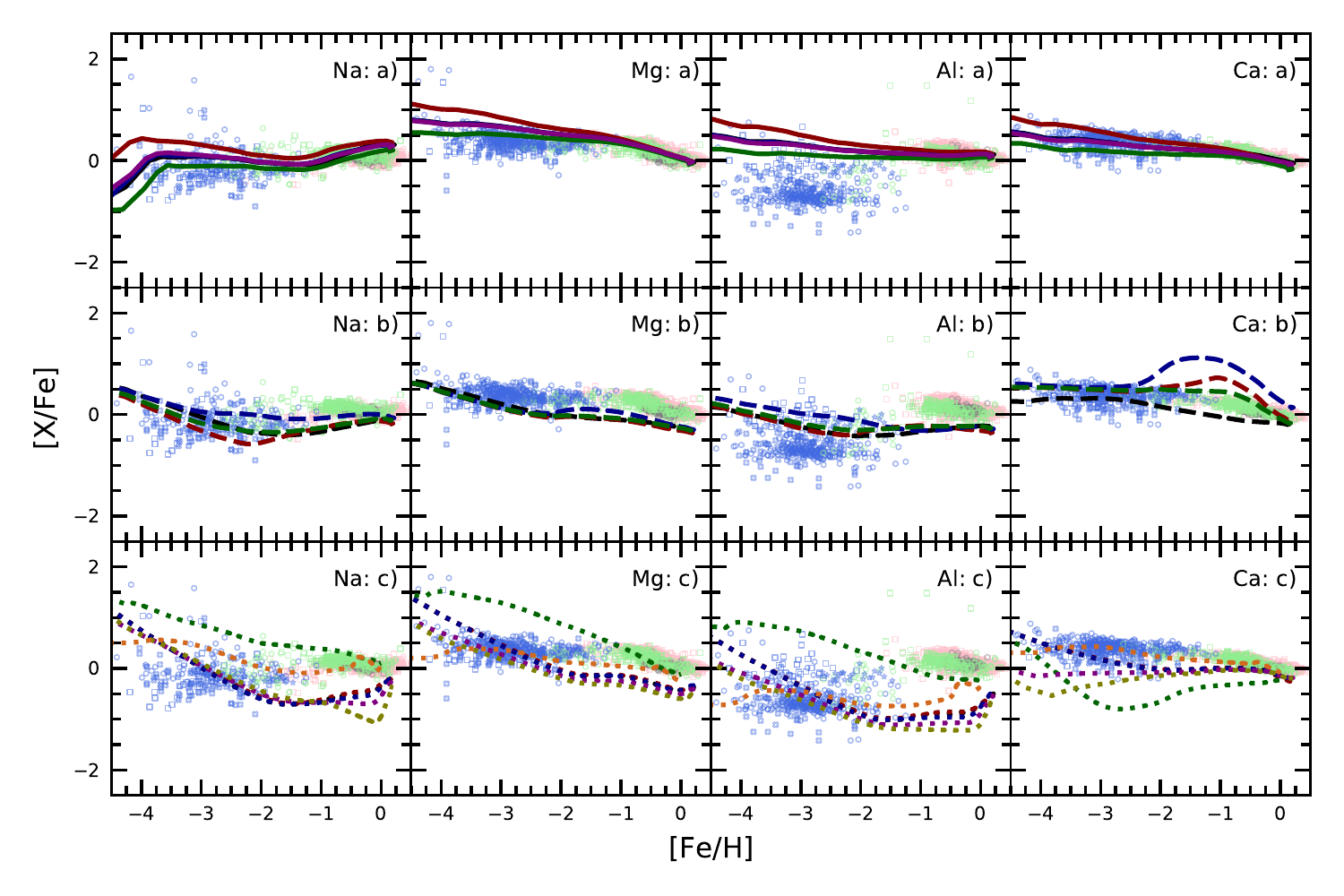}
    \caption{Similarly to Fig. \ref{fig:CNOabundance},
    the [X/Fe]-[Fe/H] relation in the Milky Way, where  each column from left to right represents Na, Mg, Al, and Ca, respectively. Each row refers to a different group. The models in the first row belong to Group A and are drawn with a solid line. The models in the middle row belong to Group B and are drawn with a dashed line. The models in the bottom row belong to Group C and are drawn with a dotted line.
    Next we describe the color patterns. \emph{first row:} Model 1 (black), Model 2 (dark red), Model 3 (dark blue), Model 4 (dark green), Model 10 (purple); \emph{second row:}  Model 5 (black), Model 6 (dark red), Model 7  (dark blue), Model 8 (dark green); \emph{third row:}  Model 9 (black), Model 11 (dark red), Model 12 (dark blue), Model 13 (dark green), Model 14 (purple), Model 15 (olive), Model 16  (chocolate). The yield prescriptions of the models are described in Table \ref{tab:yieldstabs}.
    Concerning the color patterns for the observational data:  halo stars are shown in royal blue, halo stars as well as thick and thin disc stars are shown in light green, thick and thin disc stars are shown in pink, thin disc-only stars are shown in gray. All of the data are shown with empty markers. 
    Specifically, the observational data sources are displayed as follows: . Of the halo stars in royal blue, the squares come from \citet{Cayrel04}, the star symbols come from \citet{Reggiani17}, the hexagons come from \citet{Roederer14a}, while the crosses come from \citet{Mashonkina17}. Light green halo and thin-and-thick disc stars are either circles \citep{Zhao16},  hexagons \citep{Carretta00}, or squares \citep{Reddy06}. The pink thin and thick disc stars are either circles \citep{Bensby05} or squares \citep{Bensby14}. The gray circles of thin disc stars come from \citet{Reddy03}. 
    The models are described in Table \ref{tab:yieldstabs}.}
\label{fig:NaMgAlCaabundance}
\end{figure*}

Fig. \ref{fig:CNOabundance} shows on its first column the [C/Fe]-[Fe/H] relation for data and theoretical models (Group A, B, and C in the first, second, and third row, respectively). The observational data are color-coded according to their region of sampling, as labeled in Table \ref{tab:booktabs}. We see that all models pass though the halo stars. Fig. \ref{fig:CNObox-plot} shows that in the low-metallicity regime, Group A slightly under-produces while Group B overproduces carbon with respect to the interquartile.
Both Group A \citep{Kobayashi06} and Group C \citep[NuGrid][]{NuGrid}, while managing to reach solar abundances, are unable to reproduce the slope of thin and thick disc when it comes to carbon abundances. The only exception to this trend being Model 13 of Group C, where the massive yields were computed including pre-SN nucleosynthesis and wind nucleosynthesis to the delayed model. We note however that this model  overestimates the abundances of most halo stars.  
While both Group B and C show a decreasing trend from the extremely metal-poor regime, reaching since early epochs a super-solar enrichment, Group A grows from metal-poor [C/Fe] abundances and grows in the range -4.5 dex $ \lesssim$ [Fe/H] $\lesssim$ -3.5 dex and then generally plateaus. 

\subsubsection{Nitrogen}

Nitrogen is an element with one of the richest nucleosynthetic make ups \citep[e.g.,][]{Vangioni18, Romano19}. It is produced both as a primary and as a secondary element\footnote{A primary element is synthesized from primordial gas (H and He), while a secondary element is synthesized from metals}. The production of N as a primary element was at first proposed by \cite{Truran71} and \cite{Talbot74}. Later, \cite{Matteucci86N} pointed out that in order to explain Galactic halo nitrogen abundances, also massive stars should have a primary N component. \citet{Maeder00} and \citet{Meynet02-massive} derived that if massive extremely metal-poor stars have a high initial rotational speed, they can produce primary N.  Nearly all of $^{14}$N (the most abundant nitrogen isotope) forms during hydrostatic hydrogen burning, while $^{15}$N is synthesized during both the hot and cold CNO cycles \citep[for a review, see][]{Wiescher10}. While $^{15}$N is only produced as secondary element, $^{14}$N has a primary and a secondary component.

In Fig. \ref{fig:CNOabundance} and Fig. \ref{fig:CNObox-plot}, the impact of the primary component coming from rotating low-metallicity massive stars is best exemplified by Group B. The black dashed curve in Fig. \ref{fig:CNOabundance} shows sub-solar N enrichment up to $\sim$ [Fe/H] < -2.5, coming from the secondary component produced by massive stars. With increasing metallicity, the onset of AGBs brings N up to solar abundances. All the remaining yields in Group B contain a primary N component synthesized by massive stars and hence display a plateau at low metallicities, in agreement with halo stars. Group A yields generally struggle with reproducing low-metallicity abundances, and Fig. \ref{fig:CNObox-plot} shows that models reside outside the scatter of halo stars. The models catch up to solar N abundances for disc stars, after the onset of AGB contributions. Group C models are generally inconsistent with the data, aside for Model 13, which is the model that includes pre-SN and wind nucleosynthesis to the yields of massive stars. From Fig. \ref{fig:CNOsta} it becomes clear that none of the models can quite match the super-solar and flat median of nitrogen abundances (even though Group B gets the closest), but the combination of Group A and Group B generates a standard deviation in excellent agreement with the observational data.

\subsubsection{Oxygen}

Oxygen is the better studied and better reproduced metal \citep{Matteucci12}. It is also the most abundant in nearly any astrophysical environment at scales larger than individual stars \citep[the circumstellar media of some carbon-rich AGB stars can deplete all the available oxygen into CO molecules][]{Gail09, Boyer12}. The reason why oxygen is so abundant is because it is the natural end-product of He-burning \citep{Salpeter52, Clayton03}.

Among the 12 elements considered in this work, oxygen is the one best reproduced by all yield models. Fig. \ref{fig:CNOsta} shows that  the medians of models and observational data alike are fairly consistent across metallicity bins. However, the standard deviations of the original Groups (A, B, and C) are generally small compared to the observational data values. A better performance can be seen in group combinations, especially when Group C is involved. From Fig. \ref{fig:CNOabundance} it is possible to see that Group A converges to solar oxygen abundances at solar metallicities, while the different initial rotational velocities for massive stars in Group B cause the models to span about 0.3 dex at solar metallicities.

\subsection{Alpha and odd-Z elements}\label{sec:alpha}

\begin{figure*}
\centering
    \includegraphics[width=\textwidth]{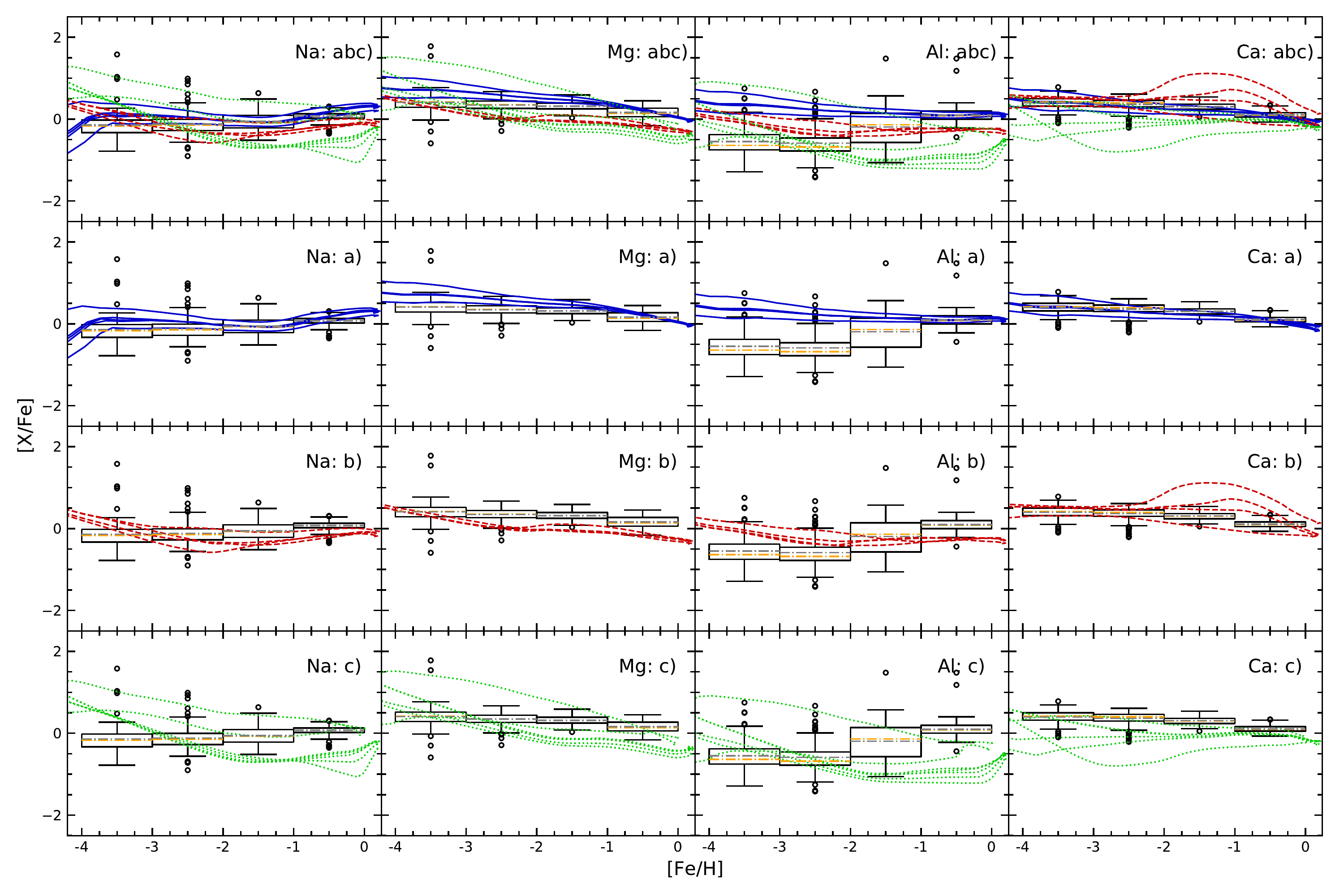}
    \caption{Similarly to Fig. \ref{fig:CNOabundance} and Fig. \ref{fig:CNObox-plot}, we show the [X/Fe]-[Fe/H] relation for the elements (Na, Mg, Al and Ca in each column from left to right) and each group, with the addition of Group ABC -- the combination of all three base groups -- on the top row. Similarly to Fig. \ref{fig:CNObox-plot}, the data are grouped in four metallicity bin in the range -4 < [Fe/H] < 0 and the box plot for each bin is shown in black. The median (dot-dashed orange lines inside the boxes) and average (dot-dashed grey lines inside the boxes) of the data are also shown. All of Group A models are shown in solid blue lines, all of Group B models are shown in dashed red lines, while all of Group C curves are shown in dotted green lines.}
\label{fig:NaMgAlCabox-plot}
\end{figure*}

\subsubsection{Sodium}
Sodium, including $^{22}$Na and $^{23}$Na, are mainly produced by hydrostatic carbon burning in massive stars dying as SNCC \citep{Woosley95,Romano10}. Other possible sources can be hydrogen envelope as consequence of the Neon-Sodium cycle and s-process on $^{22}$Ne in Helium burning \citep{Truran03,Arcones23}. As for $^{22}$Na, it is the result of proton capture on $^{21}$Ne in the carbon shell \citep{Woosley95}. Recent observations based on the Gaia-ESO Survey \citep{Gaia-ESO} indicates that the Na abundance in stars with mass below $\sim$2.0 $M_{\odot}$ seems to be unaffected by internal mixing processes. For more massive stars instead, there is an increasing trend of Na abundance with increasing stellar mass \citep{Smiljanic16}.

\begin{figure*}
\centering
    \includegraphics[width=\textwidth]{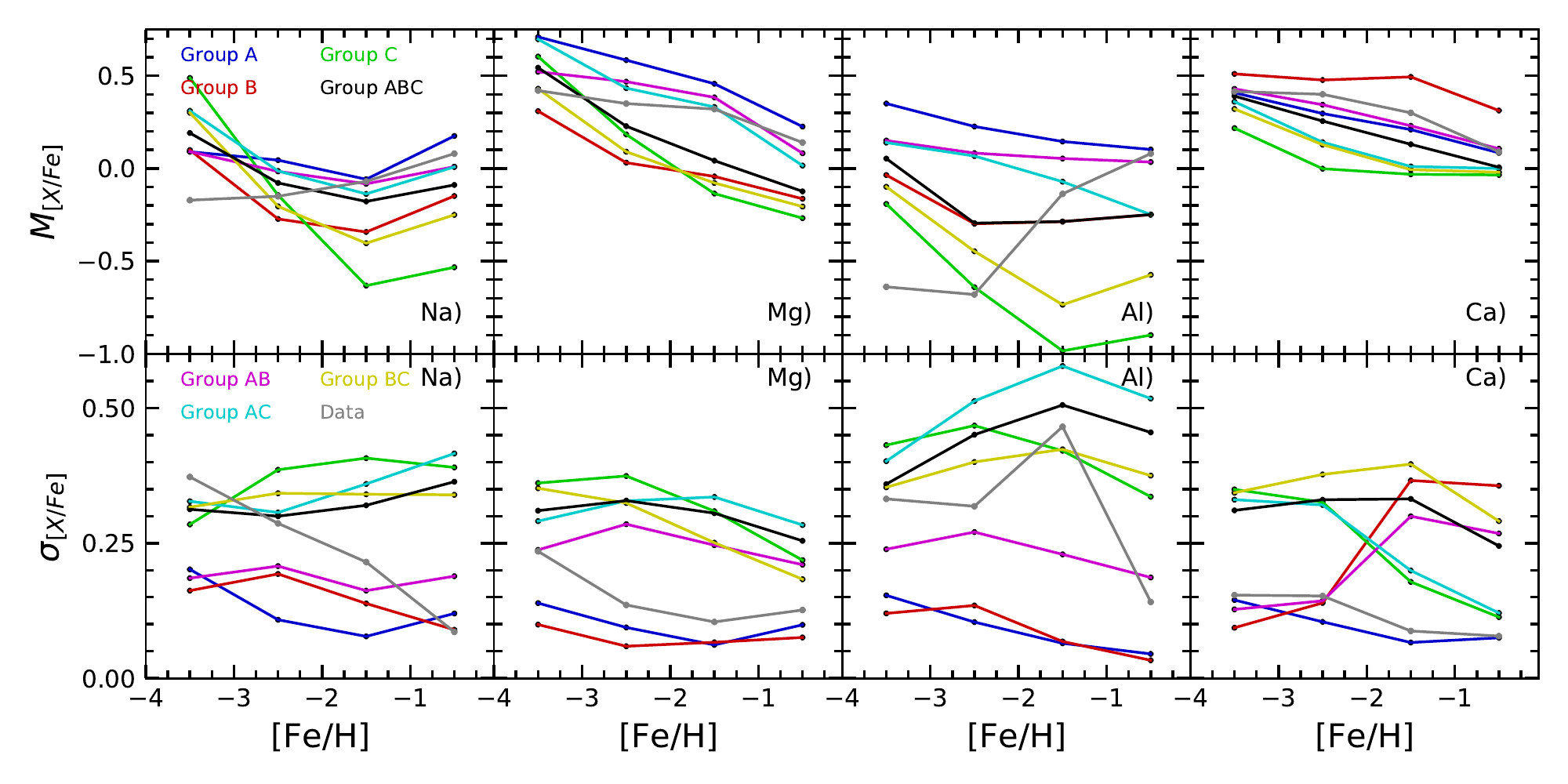}
    \caption{In metallicity bins consistent with the rest of the manuscript, we show the median ($M_{i,j,k}$, \emph{top row}) and standard deviation ($\sigma_{i,j,k}$\emph{bottom row}) for each group and group combination. 
    The black points are shown in the middle of their metallicity bins (-4.0 dex $ \lesssim$ [Fe/H] $\lesssim$ -3.0 dex, -3.0 dex $ \lesssim$ [Fe/H] $\lesssim$ -2.0 dex, -2.0 dex $ \lesssim$ [Fe/H] $\lesssim$ -1.0 dex or -1.0 dex $ \lesssim$ [Fe/H] $\lesssim$ 0.0 dex). All of the panels adopt the same colors and styles: data (grey lines), Group A (blue lines), Group B (red lines), Group C (green lines), the combination of Group A and Group B (Group AB, purple lines), the combination of Group A and Group C (Group AC, cyan lines), the combination of Group B and Group C (Group BC, yellow lines), the combination of Group A and Group B and Group C (Group ABC, black lines). A description of the statistics criteria can be found in Section \ref{sec:statistics_tool}. The data and models used are consistent with the previous figures, Fig. \ref{fig:NaMgAlCaabundance} and Fig. \ref{fig:NaMgAlCabox-plot}.}
\label{fig:NaMgAlCasta}
\end{figure*}

On the first column of Figure \ref{fig:NaMgAlCaabundance}, only Group A is in agreement with halo and disc data. Group B can fit the halo data well but underestimate the disc data  within the scatter but below the interquartile. As for Group C, all models can predict the halo data but underestimate disc data except Model 13 and Model 16. Model 13 overestimates all data while Model 16 overestimates halo data only. The scatter of data and groups for Na can be seen in Figure \ref{fig:NaMgAlCabox-plot}, first column. The halo data displays a larger scatter than disc data. However, this scatter can only be covered partially by groups except Group A at -4.0 dex $ \lesssim$ [Fe/H] $\lesssim$ -3.0 dex, -1.0 dex $ \lesssim$ [Fe/H] $\lesssim$ 0.0 dex and Group C at -2.0 dex $ \lesssim$ [Fe/H] $\lesssim$ -1.0 dex, which can be seen clearly from Figure \ref{fig:piechart1}. The median and standard deviation of data and groups can be seen in Figure \ref{fig:NaMgAlCasta}, the first column. One can see that no group can predict the increasing trend of median of data. All models can reproduce increasing trend for disc stars but only Group A or Group AB can predict similar values. Similarly, no group can predict the decreasing trend of the standard deviation for data, with the exception of Group B. However, group B cannot predict the value of the standard deviation for halo stars.

\subsubsection{Magnesium}

Magnesium, including $^{24}$Mg, $^{25}$Mg and $^{26}$Mg, exists mainly owing to hydrostatic carbon burning and explosive neon burning in massive stars \citep{Woosley95, Romano10}. Some $^{24}$Mg can be produced by silicon burning \citep{Arcones23}. As for $^{25}$Mg and $^{26}$Mg, the other two Mg isotopes, are secondary and therefore more sparsely produced. The helium burning phase reaches sufficiently high temperatures to ignite $^{22}$Ne burning into $^{25}$Mg \citep{Truran03}. Then $^{25}$Mg captures a neutron and becomes $^{26}$Mg. Nucleosynthesis of some $^{25}$Mg and $^{26}$Mg can occur in AGB stars. Similar to other $\alpha-$elements, the abundance ratio of Magnesium provides an important fossil signature in tracing different disc populations. Recent work suggests that peculiar histories of star formation are needed to reproduce  the bimodality of abundance ratio of Magnesium shown in the data \citep{Palla22}.

In the second column of Figure \ref{fig:NaMgAlCaabundance}, Group A is in agreement with disc data but overestimates halo data. Group B underestimates the disc and halo data. Group C except for model 13, which is in agreement with disc data but overestimates halo data, can fit the halo data well. The scatter of data and groups for Mg can be seen in Figure \ref{fig:NaMgAlCabox-plot}. Compared to Na, the scatter of halo data is much smaller and similar to disc data. Since the scatter of data is very small, it is hard for groups to be in agreement with data, which can be seen in Figure \ref{fig:piechart1} where only median of data can be covered by Group A at all metallicities. The median and standard deviation of data and groups can be seen in Figure \ref{fig:NaMgAlCasta}. One can see that all groups share similar decreasing trend with the median of the data. Group AB is the closest one. For the standard deviation, only Group A and Group B show a similar trend with data, but with some differences in the standard deviation value.

\subsubsection{Aluminium}
{Aluminum is mostly produced as a secondary product of hydrostatic carbon burning and neon burning \citep{Woosley95,Truran03,Romano10}. A small percentage of the $^{27}$Al is created instead in the hydrogen burning shells of evolved stars by the MgAl chain \citep{Romano10}. As for $^{26}$A1, it is synthesized hydrostatically during stellar evolution and explosively by the supernova shock in the neon or carbon burning phases \citep{Arcones23}. It is also possibly made in the hydrogen shell and dredged up by the red giant envelope or in the outer part of the helium core, which is the unburned residual of hydrogen burning \citep{Woosley95}.} {
Recent observations based on the Gaia-ESO survey indicates that Al abundance in stars with mass below $\sim$3.0 $M_{\odot}$ seem not to be overabundant while giants in cluster more massive than this are Al-rich \citep{Smiljanic16} and concludes that Al is underproduced by yields \citep{Kobayashi06,Karakas10,Ventura13-AGB-AGB,Ventura14a,Ventura14b} with large uncertainties.}

In Figure \ref{fig:NaMgAlCaabundance}, we compare the abundance data and groups for Al. There is a gap between halo data and disc data at [Fe/H] $\sim$ -1.25.  Group B and Group C can only fit the halo data while Group A can only fit the disc data. Group B and Model 13 from Group C overestimate the halo data slightly.  Figure \ref{fig:NaMgAlCabox-plot}  shows that this paucity in data causes a large scatter in the -2 < [Fe/H] < -1 metallicity bin compared to either metal-poorer or metal-richer regimes. Due to the small scatter of Group A and Group B, they cannot cover the scatter of data at all metallicities. The large Al abundances obtained in Model 13 cause a large scatter in Group C, large enough to encompass the interquartile of the data gap in the metal-poor bin. Figure \ref{fig:NaMgAlCasta}, confirms that both median and standard deviation are hardly reproduced.

\begin{figure*}
\centering
    \includegraphics[width=\textwidth]{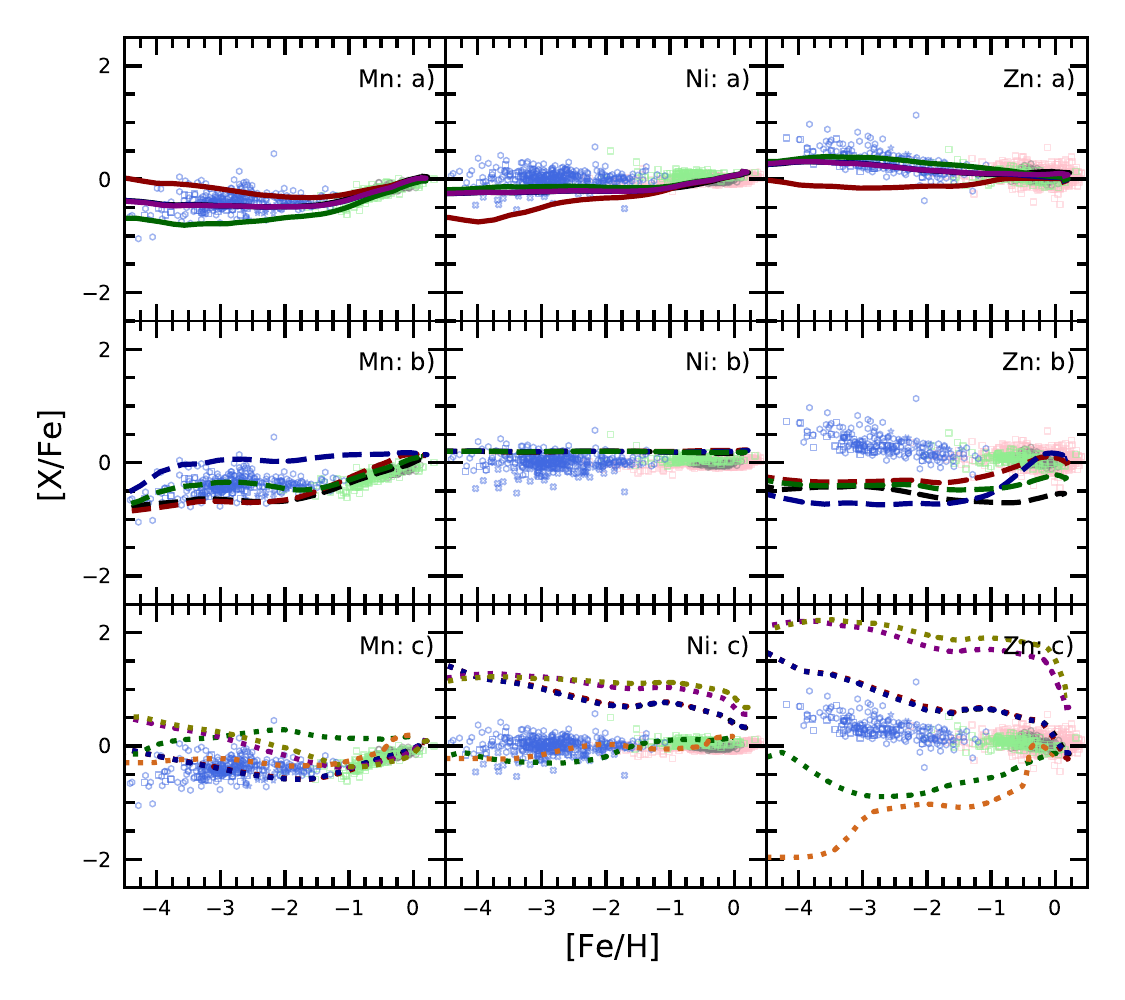}
    \caption{
    Similarly to Fig. \ref{fig:CNOabundance},
    the [X/Fe]-[Fe/H] relation in the Milky Way, where  each column from left to right represents Mn, Ni, and Zn, respectively. Each row refers to a different group. The models in the first row belong to Group A and are drawn with a solid line. The models in the middle row belong to Group B and are drawn with a dashed line. The models in the bottom row belong to Group C and are drawn with a dotted line.
    Next we describe the color patterns. \emph{first row:} Model 1 (black), Model 2 (dark red), Model 3 (dark blue), Model 4 (dark green), Model 10 (purple); \emph{second row:}  Model 5 (black), Model 6 (dark red), Model 7  (dark blue), Model 8 (dark green); \emph{third row:}  Model 9 (black), Model 11 (dark red), Model 12 (dark blue), Model 13 (dark green), Model 14 (purple), Model 15 (olive), Model 16  (chocolate). The yield prescriptions of the models are described in Table \ref{tab:yieldstabs}.
    Concerning the color patterns for the observational data: halo stars are shown in royal blue, halo stars as well as thick and thin disc stars are shown in light green, thick and thin disc stars are shown in pink, thin disc-only stars are shown in gray. All of the data are shown with empty markers. 
    Specifically, the observational data sources are displayed as follows: of the halo stars in royal blue, the squares come from \citet{Cayrel04}, the star symbols come from \citet{Reggiani17}, the hexagons come from \citet{Roederer14a}, while the crosses come from \citet{Mashonkina17}. Light green halo squares come from \citep{Reddy06}. The pink thin and thick disc stars are either circles \citep{Bensby05} or squares \citep{Bensby14}. The gray circles of thin disc stars come from \citet{Reddy03}. 
    The models are described in Table \ref{tab:yieldstabs}.
    } 
\label{fig:MnNiZnabundance}
\end{figure*}

\subsubsection{Calcium}
Calcium and its isotopes are produced by a combination of hydrostatic oxygen shell burning and explosive oxygen burning \citep{Woosley95}. It can also be made by incomplete silicon burning processes \citep{Romano10}. Both $^{40}$Ca and $^{42}$Ca exist primarily by oxygen burning \citep{Woosley95,Truran03,Arcones23}.  $^{41}$Ca is primarily the result of neutron capture by abundant $^{40}$Ca. $^{43}$Ca and $^{46}$Ca are made in the neon shells and carbon shells. Specifically, $^{44}$Ca is made almost entirely in the $\alpha$-rich freezeout, while $^{48}$Ca may be made in Type la supernova that ignite a carbon deflagration very near the Chandrasekhar mass \citep{Woosley95}.

\begin{figure*}
\centering
    \includegraphics[width=\textwidth]{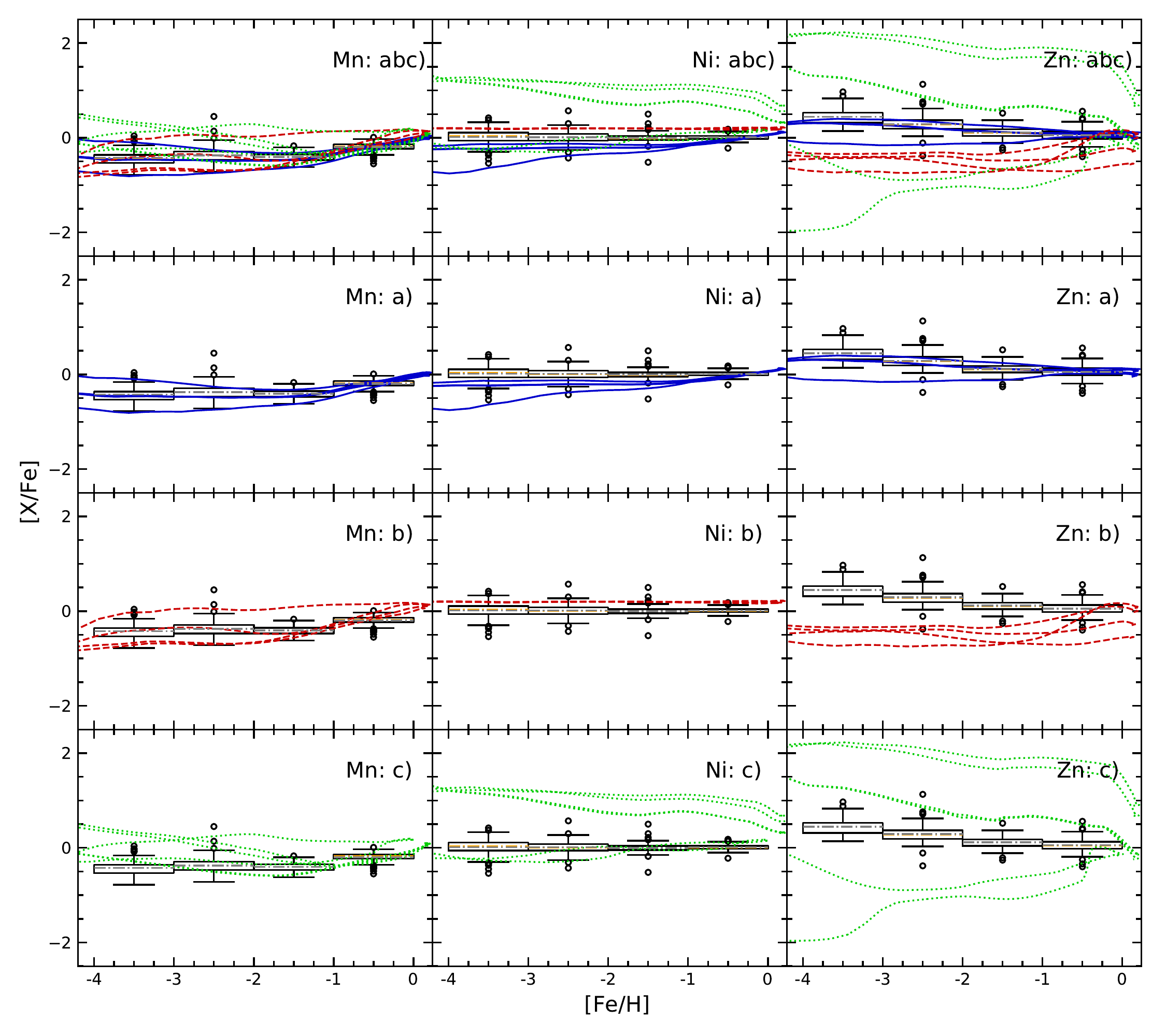}
    \caption{Similarly to Fig. \ref{fig:CNOabundance} and Fig. \ref{fig:CNObox-plot}, we show the [X/Fe]-[Fe/H] relation for the elements (Mn, Ni, and Zn in each column from left to right) and each group, with the addition of Group ABC -- the combination of all three base groups -- on the top row. Similarly to Fig. \ref{fig:CNObox-plot}, the data are grouped in four metallicity bin in the range -4 < [Fe/H] < 0 and the box plot for each bin is shown in black. The median (dot-dashed orange lines inside the boxes) and average (dot-dashed grey lines inside the boxes) of the data are also shown. All of Group A models are shown in solid blue lines, all of Group B models are shown in dashed red lines, while all of Group C curves are shown in dotted green lines.}
\label{fig:MnNiZnbox-plot}
\end{figure*}

In Figure \ref{fig:NaMgAlCaabundance} we compare observational data for Ca and groups. Group A can nicely predict the halo data and disc data, as can also be seen in Figure \ref{fig:piechart1}, while Group B overestimates disc data due to an increase at -2.0 dex $ \lesssim$ [Fe/H] $\lesssim$ -0.0 dex. Except for Model 16, Group C underestimates all data. The data scatter is very small at all metallicities in Figure \ref{fig:NaMgAlCabox-plot}.  Similar to Mg, the medians of all groups share similar trends with the median of the data, as is shown in Figure \ref{fig:NaMgAlCasta}. Group A and Group AB are in excellent agreement. When comparing the standard deviation of groups and data, it shows that only Group A can reproduce the standard deviation of data at all metallicities.

\subsection{Iron-peak elements} \label{sec:iron}

Depending on modeling assumptions such as IMF or SNIa rates, SNIa may be responsible between 50\% and 66\% of the production of isotopes around the iron peak \citep{Prantzos18, Matteucci12}.

\begin{figure*}
\centering
\includegraphics[width=\textwidth]{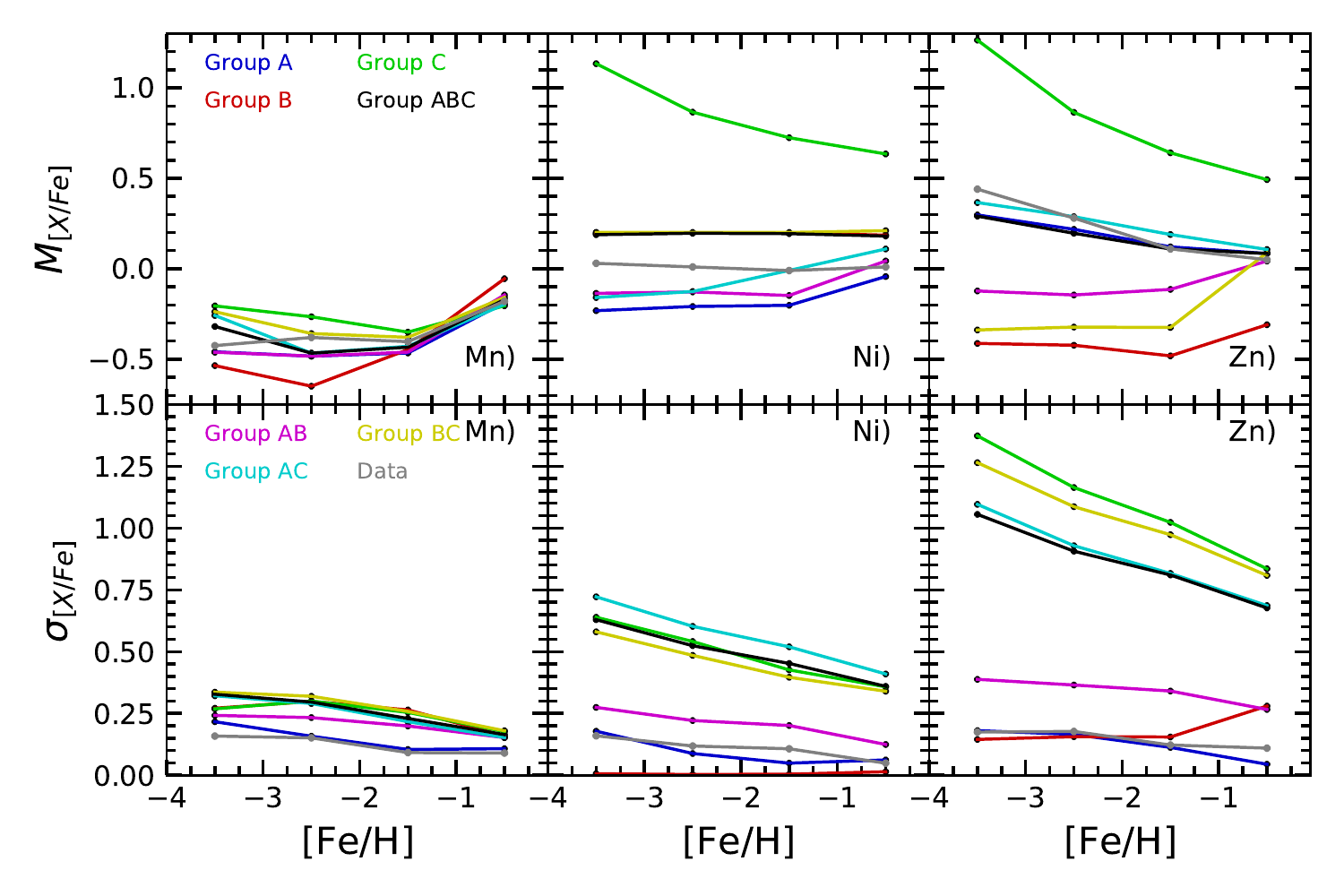}
    \caption{In metallicity bins consistent with the rest of the manuscript, we show the median ($M_{i,j,k}$, \emph{top row}) and standard deviation ($\sigma_{i,j,k}$\emph{bottom row}) for each group and group combination. 
    The black points are shown in the middle of their metallicity bins (-4.0 dex $ \lesssim$ [Fe/H] $\lesssim$ -3.0 dex, -3.0 dex $ \lesssim$ [Fe/H] $\lesssim$ -2.0 dex, -2.0 dex $ \lesssim$ [Fe/H] $\lesssim$ -1.0 dex or -1.0 dex $ \lesssim$ [Fe/H] $\lesssim$ 0.0 dex). All of the panels adopt the same colors and styles: data (grey lines), Group A (blue lines), Group B (red lines), Group C (green lines), the combination of Group A and Group B (Group AB, purple lines), the combination of Group A and Group C (Group AC, cyan lines), the combination of Group B and Group C (Group BC, yellow lines), the combination of Group A and Group B and Group C (Group ABC, black lines). A description of the statistics criteria can be found in Section \ref{sec:statistics_tool}. The data and models used are consistent with the previous figures, Fig. \ref{fig:MnNiZnabundance} and Fig. \ref{fig:MnNiZnbox-plot}.}
\label{fig:MnNiZnsta}
\end{figure*}

\subsubsection{Manganese}

Manganese has only one stable isotope, that being $^{55}$Mn. It is primarily synthesized during silicon burning and $\alpha$-rich freezeout \citep{Romano10, Nissen16}. Like the other iron-peak elements, it is an important feature of Type Ia supernovae, as shown in early spherically symmetric simulations of the carbon-deflagration model W7 \citep{Nomoto84, Thielemann86}. Manganese can also be produced  in the inner regions of explosive silicon burning in Chandrasekhar-mass deflagration models through the decay of $^{55}$Co \citep{Maeda10, Travaglio11}. Silicon burning also occurs in core collapse supernovae \citep{Thielemann96, Woosley02, Nomoto13}.  At  metal-poor regimes ([Fe/H] < -1), SNCC will dominate the Mn enrichment, while at higher metallicities where finally SNIa start occurring, both SNIa and SNCC yield roughly equal contributions of Mn. This increasing trend can be seen in all models in Figure \ref{fig:MnNiZnabundance} and Figure \ref{fig:MnNiZnbox-plot}, except for Model 7 and Model 13. Model 7 assumes the highest initial rotational velocity for massive stars (300 km/s) for LC18, which is unrealistic given that such conditions are more likely to occur at high metallicities only.  For roughly all models, the standard deviation is small, but larger than the uncertainty of the data (Figure \ref{fig:MnNiZnbox-plot} and \ref{fig:MnNiZnsta}). As for the median, with the exception of Group B, all models display a near-plateau in the metal-poor regions, and then an increase with the onset of SNIa. 

\subsubsection{Nickel}

Nickel comes in a variety of isotopes from an atomic mass of 56 to 64. Species 56 and 57 are radioactive $\gamma$-ray emitters in supernovae. These Ni isotopes rapidly decay to their Fe isobars. $^{58}$Ni is the most abundant Ni isotope, accounting for nearly 2/3 of all Ni species \citep{Clayton03}
. Ni species are synthesized like the other iron-peak elements during the explosive silicon burning and alpha-rich freezeout in supernovae. Both SNIa and SNCC produce Ni, in roughly equal parts \citep[e.g., ][]{Arcones23, Matteucci12, Andrews17}.

In Figure \ref{fig:MnNiZnabundance}, the yields of Group B have this striking flat [Ni/Fe] feature. \citet{Prantzos18} who uses the same yield set \citep{Limongi18} for massive stars, obtains a moderate rise of [Ni/Fe] with metallicity. They attribute the rise to the known overproduction of Ni in the W7 models \citep[e.g.][as seen in Figure \ref{fig:SNIayields}]{Iwamoto99}. Our fiducial SNIa yield is \citet{Seitenzahl12}, which indeed yields less Ni compared to W7.

\subsubsection{Zinc}

Zinc in synthesized both as an s-process element  and through the alpha-rich freezeout of SNCC \citep{Frohlich06}. The latter requires very high electron fractions with an excess of 0.5 in the layers undergoing Si burning. The standard deviation of Zn stellar abundances is quite small (Figure \ref{fig:MnNiZnbox-plot} and Figure \ref{fig:MnNiZnsta}) and a modest decreasing trend toward solar values with increasing metallicity can be appreciated in the data.
Figure \ref{fig:MnNiZnabundance} shows that while Group A seem to be in good agreement with the data, Group B and Group C fail at reproducing the trends. Group B consistently underproduces Zn, while Group C both wildly overproduces and underproduces Zn 

\subsection{Neutron-capture elements}\label{sec:ncapture}

Isotopes beyond the iron peak are primarily synthesized through neutron capture. An isotope inundated by a neutron flux will have a  probability to capture a neutron which depends on the neutron density in a flux \citep[detailed calculations first proposed by][]{Burbidge57}. The astrophysical sites where neutron capture occurs are predominantly clustered around two peaks: one originates from low neutron-density fluxes, so that the neutron capture occurs slowly with respect to its beta decay (s-process) -- and the other originates from  catastrophic events whose fluxes are very neutron rich. In the latter case, the neutron capture occurs rapidly (r-process) relative to the isotope's beta decay timescale. The two principal sites of s-process enrichment are the hydrostatic core He-burning and shell C-burning of both LIMs during their AGB phase and of massive stars \citep{Prantzos20}. There is yet no definitive evidence on which are the dominant sites for r-process enrichment. NSM have been the leading candidates for decades \citep{Lattimer74, Thielemann17}, and the gravitational wave event 
GW170817 \citep{Abbott17} and the associated kilonova AT2017gfo \citep{Smartt17} confirmed for the first time, as evidenced by the direct observation of strontium in day-1 spectra \citep{Watson19} that NSM do indeed produce r-process elements. {However, despite their potential as a significant r-process source, NSM alone have repeatedly been shown to struggle in explaining the high r-process enrichment of the most metal-poor stars, even when dynamics modeling is included \citep{Argast04, vandeVoort20}. Nevertheless, some studies have argued the contrary \citep[e.g.,][]{Shen15}. Within hydrodynamical simulations, \citet{Haynes19} suggest that NSM in conjunction with magneto-hydrodynamical supernovae could describe the abundance of r-process elements, as exemplified by europium abundances. Yet the authors acknowledge that NSM alone are insufficient, a result later confirmed by \citet{Kobayashi23}. }

\begin{figure}
\centering
\includegraphics[width=\columnwidth]{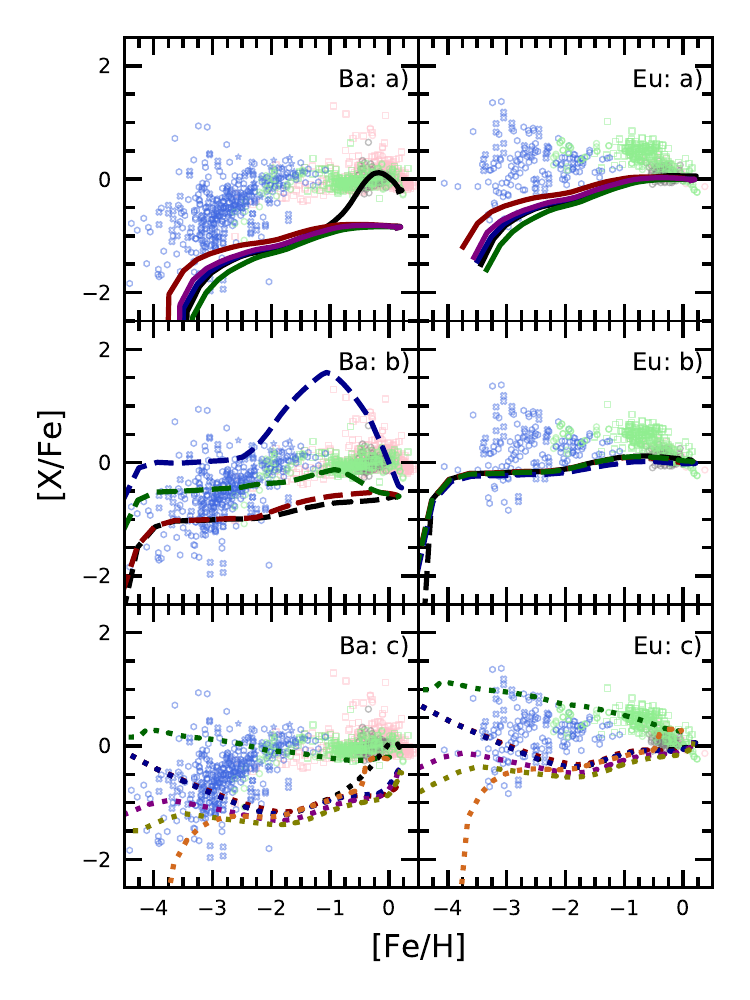}
    \caption{The [X/Fe]-[Fe/H] relation in the Milky Way, where  each column from left to right represents Ba and Eu, respectively. The color patterns for observational data: data contained halo stars only are colored royal blue,  data contained thin disc only are colored gray, data contained both thick and thin disc are colored pink, data contained halo, thick and thin disc are colored light green. All of them are empty markers.
    Specifically, the observational data sources  are labeled as follows: pink empty circles \citet{Bensby05}; pink empty squares \citet{Bensby14}; gray empty circles \citet{Reddy03}; light green empty squares \citet{Reddy06}; royal blue empty hexagons \citet{Roederer14a}; light green empty circles \citet{Zhao16}; royal blue empty stars \citet{Reggiani17}; royal blue empty crosses \citet{Mashonkina17}. The models in the first row belong to Group A and are drawn with a solid line. The models in the middle row belong to Group B and are drawn with a dashed line. The models in the bottom row belong to Group C and are drawn with a dotted lne. 
    Next we describe the color patterns.    \emph{first row:} Model 1 (black), Model 2 (dark red), Model 3 (dark blue), Model 4 (dark green), Model 10 (purple); \emph{second row:}  Model 5 (black), Model 6 (dark red), Model 7  (dark blue), Model 8 (dark green); \emph{third row:}  Model 9 (black), Model 11 (dark red), Model 12 (dark blue), Model 13 (dark green), Model 14 (purple), Model 15 (olive), Model 16  (chocolate). The models are described in Table \ref{tab:yieldstabs}} 
\label{fig:BaEuabundance}
\end{figure}

\begin{figure}
\centering
    \includegraphics[width=\columnwidth]{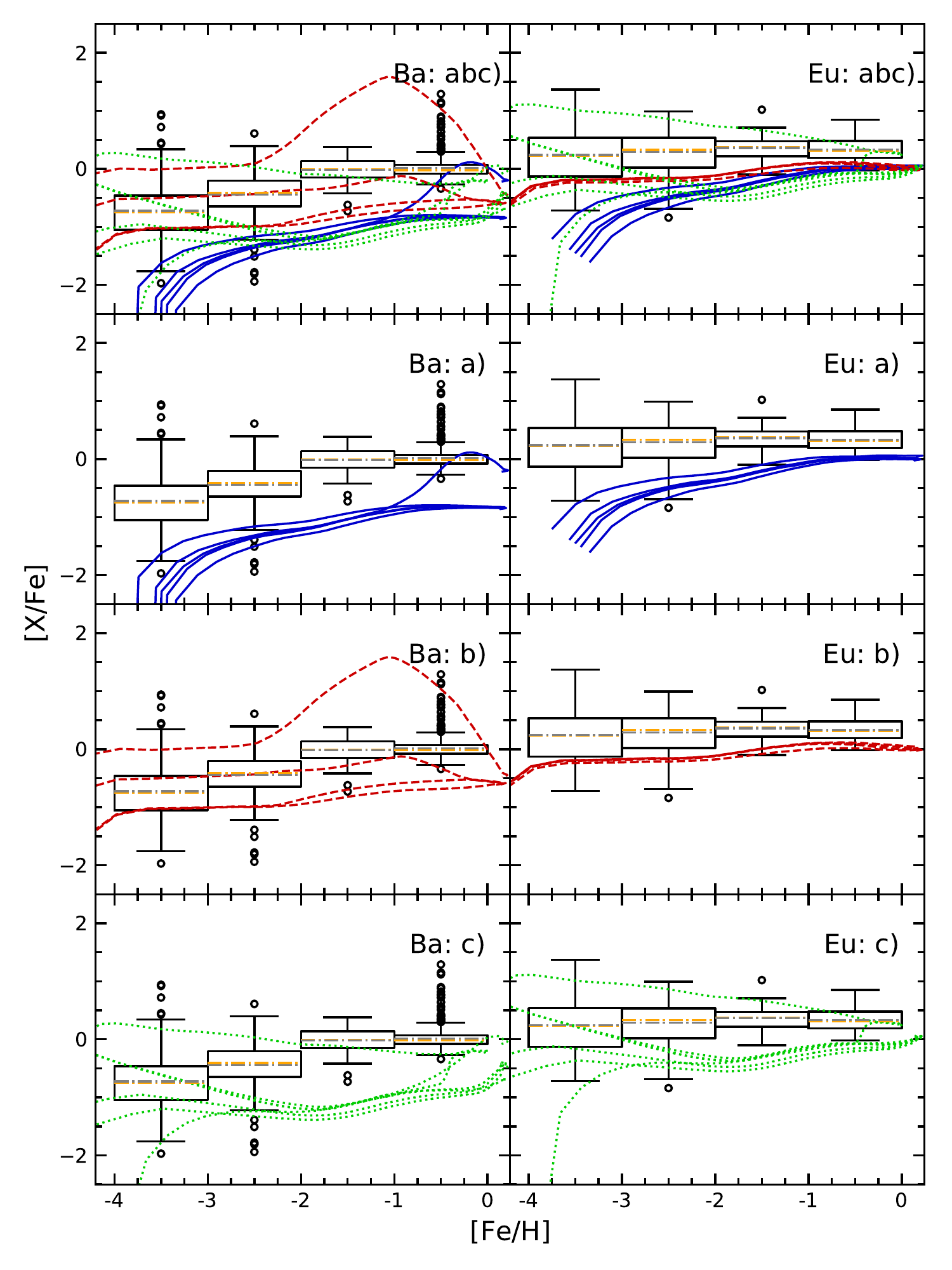}
    \caption{Similarly to the previous box-plots, we show the [X/Fe]-[Fe/H] relation for the elements (Ba and Eu in each column from left to right) and each group, with the addition of Group ABC -- the combination of all three base groups -- on the top row. Similarly to Fig. \ref{fig:CNObox-plot}, the data are grouped in four metallicity bin in the range -4 < [Fe/H] < 0 and the box plot for each bin is shown in black. The median (dot-dashed orange lines inside the boxes) and average (dot-dashed grey lines inside the boxes) of the data are also shown. All of Group A models are shown in solid blue lines, all of Group B models are shown in dashed red lines, while all of Group C curves are shown in dotted green lines.}
\label{fig:BaEubox-plot}
\end{figure}

\subsubsection{Barium}
Both the s-process and the r-process yield a significant amount of  Ba. The s-process, occurring at later stages of galaxy evolution with the onset of AGB enrichment, cannot account for the enrichment at low metallicities (below [Fe/H] $\lesssim$ -2.5). \citet{Pagel97} proposed that at the earliest stages, Ba may be produced as an r-process element. The authors suggested that massive stars may be the source of r-process enrichment. The necessity for a fast enrichment component was confirmed in detailed GCE studies \citep{Cescutti06}. However, the standard evolution of massive stars cannot reach the r-process regimes, therefore other processes associated with massive stars such as magnetorotational hydrodynamic jets or collapsars could be at play \citep{Limongi18, Siegel19, Shibagaki16}. The onset of NSM is delayed, therefore this mechanism also cannot be responsible for early Ba enrichment \citep{Kobayashi20, Yamazaki22}.

As seen in Figure \ref{fig:BaEuabundance}, in the halo, or otherwise at very metal-poor metallicities, the abundance of Ba is low, ranging from [Ba/Fe] = -2 to moderately super solar abundances.  In Group A, the variation among the models is caused by the slight difference in iron production over time. None of the models in Group A contain s-process enrichment, with the exception of Model 1. The steep increase at high metallicities of the black curve is caused by the AGB s-process enrichment from the F.R.U.I.T.Y. yields. Group B and group C models also include the s-process. If rapidly rotating massive stars (Model 7 in dark blue of Group B) were to occur at high metallicities, they also could cause a dramatic increase in Ba abundances through the s-process. The models in all groups include r-process enrichment through NSM.

\begin{figure}
\centering
    \includegraphics[width=\columnwidth]{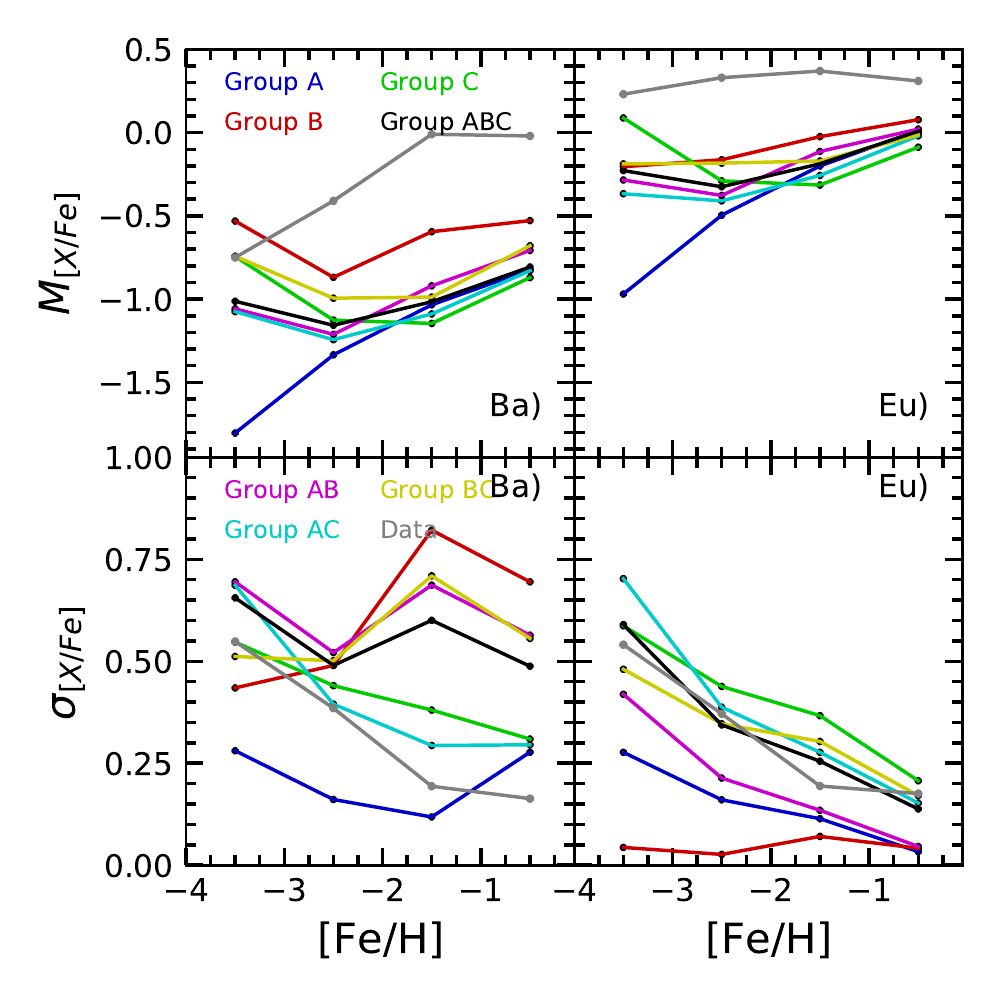}
    \caption{In metallicity bins consistent with the rest of the manuscript, we show the median ($M_{i,j,k}$, \emph{top row}) and standard deviation ($\sigma_{i,j,k}$\emph{bottom row}) for each group and group combination. 
    The black points are shown in the middle of their metallicity bins (-4.0 dex $ \lesssim$ [Fe/H] $\lesssim$ -3.0 dex, -3.0 dex $ \lesssim$ [Fe/H] $\lesssim$ -2.0 dex, -2.0 dex $ \lesssim$ [Fe/H] $\lesssim$ -1.0 dex or -1.0 dex $ \lesssim$ [Fe/H] $\lesssim$ 0.0 dex). All of the panels adopt the same colors and styles: data (grey lines), Group A (blue lines), Group B (red lines), Group C (green lines), the combination of Group A and Group B (Group AB, purple lines), the combination of Group A and Group C (Group AC, cyan lines), the combination of Group B and Group C (Group BC, yellow lines), the combination of Group A and Group B and Group C (Group ABC, black lines). A description of the statistics criteria can be found in Section \ref{sec:statistics_tool}. The data and models used are consistent with the previous figures, Fig. \ref{fig:BaEuabundance} and Fig. \ref{fig:BaEubox-plot}.}
\label{fig:BaEusta}
\end{figure}

Figure \ref{fig:BaEubox-plot} the subsolar interquartile range  shows clearly that amid many outliers, the standard deviation of Ba is large at low metallicities and small at high metalliciites. None of the models display a particularly good fit to the data, with the exception of Model 8 in Group B, i.e. the set averaged across the models of varying rotational velocities. Figure \ref{fig:BaEusta} shows very poor agreement between models and data. All models fall below the median values in nearly all metallicity bins, and the scatter trends show generally little correlation.

\subsubsection{Europium}

The two stable isotopes of Eu, 151 and 153, are produced exclusively through r-process enrichment \citep{Prantzos20, Cowan04}. As such, paired with Ba, Eu is often used as a discriminating indicator on whether heavy elements in a star originate predominantly from s- or r-process enrichment \citep[e.g.,][]{Aguado21}. Subsolar [Ba/Eu] abundances would indicate a predominance of r-process enrichment while supersolar [Ba/Eu] abundances would instead favor s-process enrichment.

Figure \ref{fig:BaEuabundance} and \ref{fig:BaEubox-plot} show that the models barely reach the lower data scatter for Group A, or the 25th percentile for group B. Similar underproduction can also be observed for Group C, with the exception of Model 13, where the pre-SN nucleosynthesis and wind nucleosynthesis are added to the delayed model. Given that none of the massive yields in our models produce r-process enrichment, the abundance trends are exclusively dictated by the addition of NSM onto the base GCE models. 
These results calibrated on a Milky-Way-like galaxy are in agreement with literature results in that NSM occur too late in the chemical enrichment history to explain the wide spread of the abundances for Eu enhancement, especially at low metallicities.

\subsection{Present-day abundance ratios}
\begin{figure}
\centering
\includegraphics[width=\columnwidth]{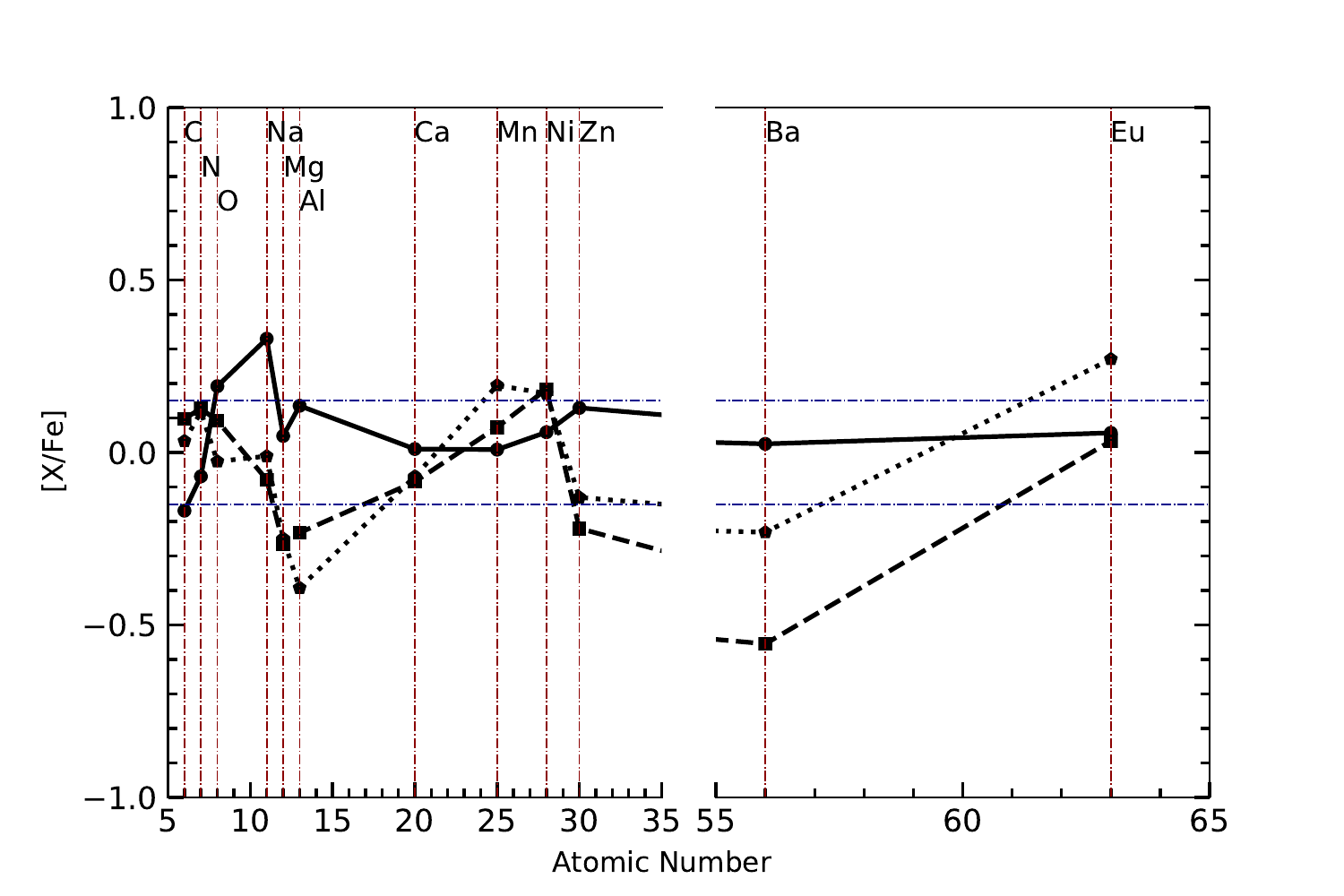}
\caption{Abundance ratios over iron of the 12 elements investigated in the present work normalized to solar values. The abundances are taken wherever the model reaches [Fe/H] = [Fe/H]$_{\odot}$. The x-axis indicates the atomic number. The elements are highlighted by the vertical (dark-red dot-dashed) lines. The black curves refer to the various models. We choose one representative model from each group: Model 1 from Group A (solid line), Model 8 from Group B (dashed line) and Model 16 from Group C (dotted line). The horizontal dark blue dot-dashed lines mark a $\pm$ 0.15 dex deviation from the solar normalization.}
\label{fig:X_vs_atomicnumber}
\end{figure}

 {In this last result presented in Fig. \ref{fig:X_vs_atomicnumber}, we show the abundance produced by 3 models selected from 3 different groups (Model 1, 8, 16) for all of the 12 elements. The abundances have been normalized to solar values. The horizontal lines identify a range of $\pm$ 0.15 dex. These 3 models show the best performance in their corresponding group. Model 1 underestimates Carbon abundance and overestimates Oxygen and Sodium abundance. As for other elements, the difference between reproduced abundance and solar abundance is less than 0.15 dex. The other 2 models (Group B and Group C) underestimate the abundance of Magnesium, Aluminium and Barium. Model 8 (Group B) overestimates Nickel while Model 16 (Group C) overestimates Manganese, Nickel and Europium.}

\section{Discussion and Conclusions}\label{sec:Conclusion}

\begin{figure*}
\centering
\includegraphics[width=\textwidth]{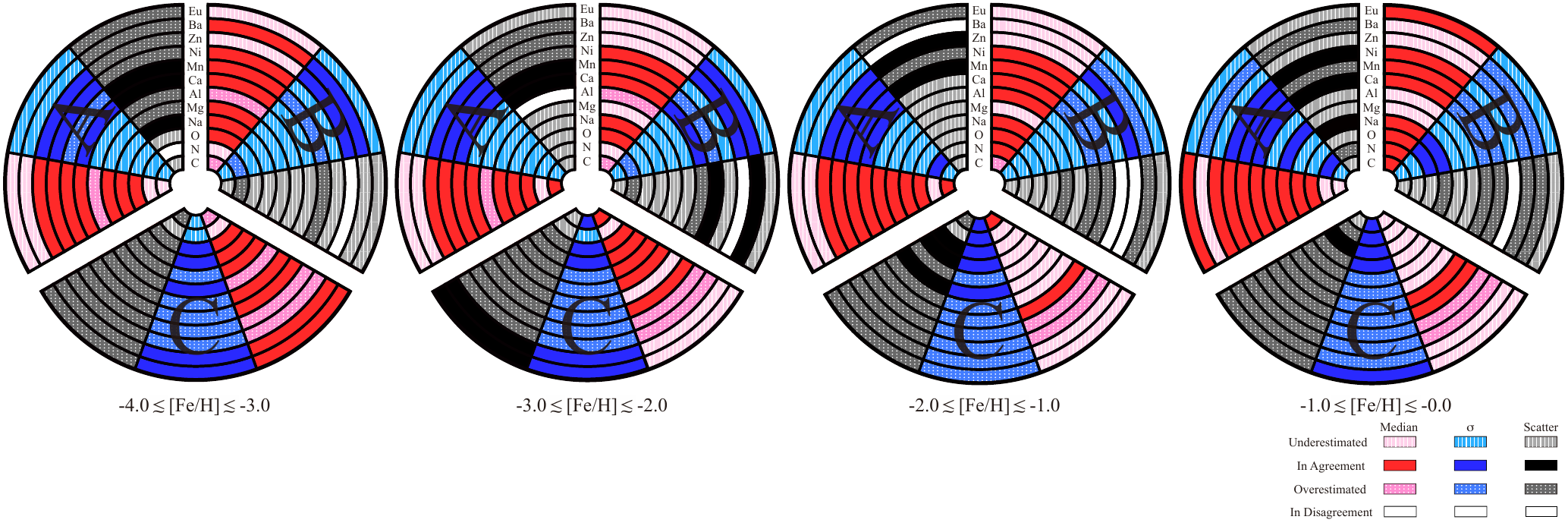}
\caption{For each of Group A, B, and C, the summary of the Judgement criteria (Median, standard deviation, and scatter, Section \ref{sec:statistics_tool}) for the four metallicity bins in the range -4 < [Fe/H] < 0. Each pie chart is displayed in increasing metallicity from left to right. Each shell refers to one of the 12 elements investigated in this work. From the inner to the outer shell, we show carbon, nitrogen, oxygen, sodium, magnesium, aluminum, calcium, manganese, nickel, zinc, barium and europium. The left slice in each pie chart depicts Group A, the right slice Group B, while the bottom slice Group C. Median, standard deviation, and scatter are shown in red, blue, and black, respectively. 
If the shell slice has a solid color, it means that the Judgement for the statistical quantity for the observed abundances is in agreement with the models. If the shell slice is slightly fainter with a dotted texture, it means that the data are overestimated by the models. For the faintest shell slices with a striped texture, the data are underestimated by the model.}
\label{fig:piechart1}
\end{figure*}

\begin{figure*}
\centering
\includegraphics[width=\textwidth]{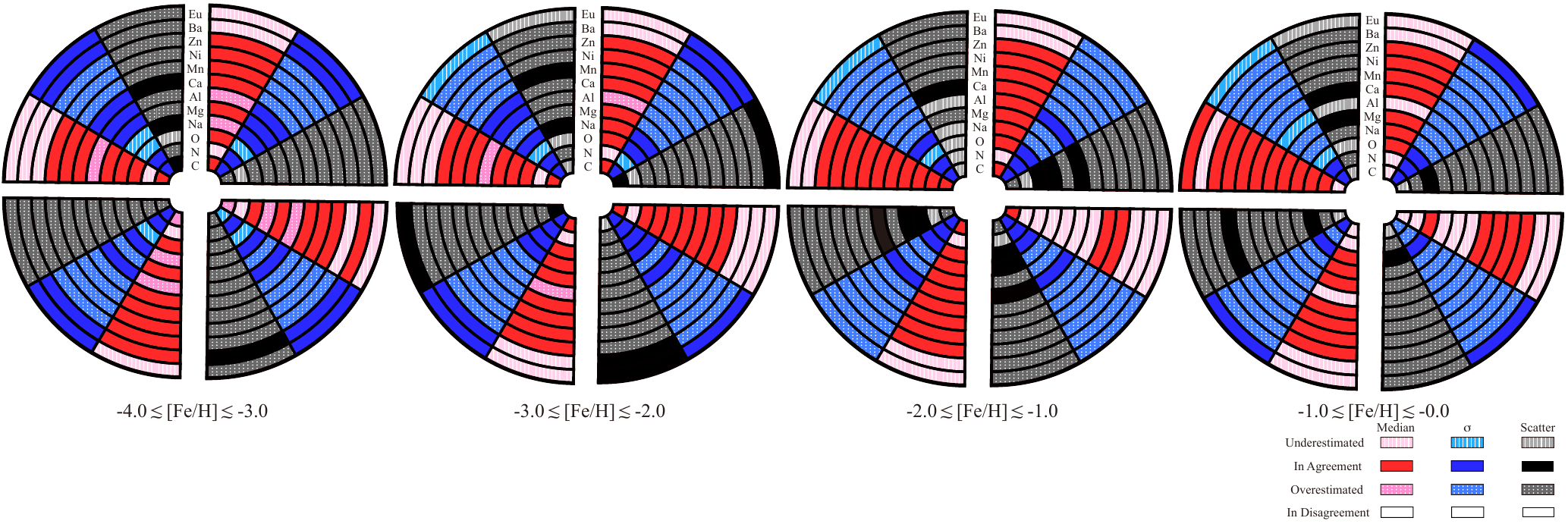}
\caption{Similarly to Fig. \ref{fig:piechart1}, we show the pie charts for group combinations: Group AB is represented in the top left slice, Group AC in the top right slice, Group BC in the bottom right slice, and all the models are combined in the bottom left slice with Group ABC.}
\label{fig:piechart2}
\end{figure*}

In this paper we investigated the impact of the choice of yield tabulations on chemical evolution runs produced by a single NuPyCEE parametrization, fine-tuned to the Milky Way. We compared the model variations with observational data coming from the major Galactic environments: thin disc, thick disc, and halo -- and combinations thereof. The summary of statistics for groups and their combinations for all elements compared with the stellar abundance data sorted in the same metallicity bins are shown in Fig. \ref{fig:piechart1} and Fig. \ref{fig:piechart2}. {The yield variations in our study are divided into three groups: Group A, which incorporates different hypernovae fractions; Group B, which varies the initial rotational velocity of massive stars; and Group C, which explores variations in the delayed/rapid explosion prescriptions for massive stars. } The criteria of our judgements for both figures are defined in Sec. \ref{sec:statistics_tool}.

Our main conclusions  based on Fig. \ref{fig:piechart1} and Fig. \ref{fig:piechart2} can be summarized as follows:

\begin{enumerate}
    \item {The medians for single groups as well as group combinations is in good agreement with the Galactic observational data for most elements up until Zinc. However, neither the scatter nor the standard deviation of groups and group combinations tend to correlate with observational trends.} Specifically, Group A and Group B underestimate most of the standard deviation and scatter of most of the data while Group C overestimates them. As for the median, Group A can predict nearly all of the data at all metallicities while Group B can predict about half of the data at all metallicities. Group C can only predict about half of the data at low metallicities and underestimates most of the data at -2.0 dex $ \lesssim$ [Fe/H] $\lesssim$ 0.0 dex.
    \begin{itemize}
    \item[$\bullet$] {Combining 2 or 3 groups at a time}, the underestimated standard deviation and scatter of data only occupy a relatively small fraction at all metallicities. Although these combinations increase the fraction of the reproduced data scatter and reproduced standard deviation at low metallicities, they also increase the fraction of overestimated scatter and standard deviation. {Standard deviation and scatter of data tend to be overestimated in the range -2.0 dex $ \lesssim$ [Fe/H] $\lesssim$ 0.0 dex}.
    \end{itemize}

    \, 
    
    \item {We draw the following conclusions regarding the breakdown of elemental groups:}
    \begin{itemize}
    \item[$\bullet$] For CNO elements, Group A and Group B underestimate most of the standard deviations of these 3 elements while Group C is in agreement with all of the standard deviations at -2.0 dex $ \lesssim$ [Fe/H] $\lesssim$ 0.0 dex. {Regarding the scatter, it appears that the majority of the estimates are too small for all three groups.}. Most of the standard deviation of these 3 elements can be reproduced by group combinations at all metallicities.
    \item[$\bullet$] {In the case of Odd-Z elements (Na and Al in our sample), we found that the median value of Na can be accurately reproduced by two of the groups. Group C performs well at lower metallicities, while it underestimates Na at -2.0 dex $\lesssim$ [Fe/H] $\lesssim$ 0.0 dex. However, Group C is in agreement with the standard deviation and most of the scatter for Na, whereas the other two groups tend to underestimate these values. In contrast to Na, Al does not exhibit clear trends.}
    \item[$\bullet$] In the case of $\alpha$ elements in our sample (Mg and Ca in our sample), we found that Group A and Group B are able to reproduce most of the standard deviation and scatter of Ca, while Group C tends to overestimate these values. However, when it comes to the median of Ca, Group A and Group B are in good agreement at all metallicities, while Group C only agrees with the median at low metallicities. For Mg, both Group A and Group B tend to underestimate the standard deviation and scatter, while Group C still overestimates them. The trend for medians is similar for both Ca and Mg. In general, the median values of both Ca and Mg can be accurately reproduced for any combination of groups, although the majority of the standard deviation and scatter tend to be overestimated.
    \item[$\bullet$] Our analysis of iron-peak elements shows that Group A is able to accurately reproduce most of the median and standard deviation values of these elements at all metallicities, while Group C tends to overestimate them, as shown in Fig. \ref{fig:piechart1}. Group B is able to accurately reproduce the median values of these elements, with the exception of Zn, but can be in agreement with the standard deviation of Zn at -4.0 dex $\lesssim$ [Fe/H] $\lesssim$ -1.0 dex. When combining the three groups, the median values of iron-peak elements can be accurately reproduced at all metallicities, with the exception of a few data points for Zn at low metallicities. However, all of the standard deviation values for these elements tend to be overestimated by the combined groups, as shown in Fig. \ref{fig:piechart2}. It's worth noting that iron-peak elements can be reproduced more accurately by the combination of the three groups than other elements, which are primarily reproduced by Group A.
    \item[$\bullet$] Neutron-capture elements (Ba and Eu), exhibit the most significant tension between observational data and theoretical models. Either single group or groups combination underestimate most of the median of Eu and Ba. As for the standard deviation, Group B and group combinations can be in agreement with most of them at low metallicities. At -2.0 dex $ \lesssim$ [Fe/H] $\lesssim$ 0.0 dex, single group and groups combinations tend to overestimate Ba and underestimate Eu. As for scatter, either single group or group combination overestimate most of them.
    \end{itemize}
\end{enumerate}

{
This paper introduces statistical criteria that compare the median, standard deviation, and scatter of observational data with different groups of models, providing a measure of the ability of each yield combination to match observational data. By analyzing the effectiveness of individual yields in describing average Galactic patterns, we highlight the impact of various yield modeling assumptions on theoretical tracks. Our work shows that the abundance scatter patterns in the Milky Way cannot be explained solely by the simultaneous application of the proposed yield prescriptions.

Our objective in this analysis was to investigate whether multiple yield prescriptions and their respective scenarios could be simultaneously at play in the evolution of our Galaxy and partially responsible for the scatter of the data. We proposed statistical criteria that compares the median, standard deviation, and scatter of observational data against groups of models to evaluate the ability of each yield combination to match observational data. However, our proposed method is inconclusive due to several factors. First, a realistic mixing of yield prescriptions cannot be described by a simple average of the models without taking into account weights or time/metallicity dependence. Introducing additional modeling parameters and assumptions to describe the variation of the occurrence of each yield prescription over time would be necessary, but this goes beyond the scope of our analysis. Second, the average of evolutionary tracks is different from a single evolutionary track containing an average or mixing of the yields, with the latter being more realistic. 

One-zone GCE models are appropriate for describing global evolutionary trends of elemental abundances. However, in order to analyze the spread in observational data, it is insufficient to only vary yields on a fixed galactic environment. The source of the spread in the data is likely due to varying physical or dynamical properties of the local environment, rather than the simultaneous involvement of multiple nucleosynthetic mechanisms.

While many yield prescriptions are generally effective at reproducing median Galactic trends, the inconsistency of scatter and standard deviation leads us to conclude that yield mixing may not be as significant a factor in the observed scatter of the data compared to physical properties or dynamical effects. At the same time, we cannot rule out that multiple yield prescriptions could be occurring, because several of these prescriptions are equally effective in describing average trends. To further investigate our objectives, improvements can be achieved through the use of grids of GCE models spanning a range of physical parameters, multi-zone GCE models, or hydrodynamical simulations.

Out of all the elements analyzed in this study, neutron-capture elements are the least accurately reproduced. Our analysis only accounts for merging neutron stars as an r-process enrichment channel. We therefore highlight the importance of exploring alternative production channels, such as collapsars and magneto-hydrodynamic jets. 

}

\section*{Acknowledgements}

    We thank the anonymous referee for the meticulous feedback.
    We thank Benoit C{\^o}t{\'e} for providing the complete Milky Way NuPyCEE parametrization of his 2019 work \citep{Cote19} and for fruitful discussions.
    E.G. acknowledges the support of the National Natural Science Foundation of China (NSFC) under grants NOs. 12173016, 12041305. E.G. acknowledges the science research grants from the China Manned Space Project with NOs. CMS-CSST-2021-A08, CMS-CSST-2021-A07. E.G. acknowledges the Program for Innovative Talents, Entrepreneur in Jiangsu.
    All the authors have been supported by the National Natural Science Foundation of China under grant (No.11922303) and the Fundamental Research Funds for the Central Universities (No.2042022kf1182).

\section*{Data Availability}

No new data were generated or analysed in support of this research.



\bibliographystyle{mnras}
\bibliography{Jinningbibliography} 





\bsp	
\label{lastpage}
\end{document}